\documentclass[12pt,a4paper,reqno]{amsart} 

\usepackage{amssymb,amsthm}
\usepackage{amsmath}
\usepackage{amsaddr}
\usepackage{graphicx}
\usepackage[colorlinks,hyperfootnotes=false]{hyperref}
\hypersetup{pdfauthor={Nathan Haouzi and Christian Schmid},pdftitle={Little String Defects and Bala-Carter Theory}, pdfkeywords={Surface Defects, Bala Carter, Little Strings}, pdfdisplaydoctitle={true},bookmarksopen={true}} 
\usepackage[all]{hypcap}	
\usepackage{url}
\usepackage{epstopdf}
\usepackage{subcaption}
\usepackage{booktabs}
\usepackage{dsfont} 
\usepackage{tikz}
\usepackage{ytableau} 
\usepackage{mciteplus} 
\usepackage{setspace}
\usepackage{xcolor}
\usepackage{colortbl}
\usepackage[skins]{tcolorbox}
\usepackage{longtable}
\usepackage{array}
\usepackage{pdflscape} 
\usepackage{placeins}

\usetikzlibrary{calc}
\usetikzlibrary{decorations.pathreplacing}
\usetikzlibrary{shapes.misc}

\newcommand{\beq}{\begin{equation}}
\newcommand{\eeq}{\end{equation}}

\newcommand{\fb}{\mathfrak{b}}
\newcommand{\fg}{\mathfrak{g}}
\newcommand{\fh}{\mathfrak{h}}
\newcommand{\fl}{\mathfrak{l}}
\newcommand{\fm}{\mathfrak{m}}
\newcommand{\fn}{\mathfrak{n}}
\newcommand{\fp}{\mathfrak{p}}

\newcommand{\fu}{\mathfrak{u}}

\newcommand{\cC}{\mathcal{C}}
\newcommand{\cN}{\mathcal{N}}
\newcommand{\cW}{\mathcal{W}}
\newcommand{\cO}{\mathcal{O}}

\newcommand{\cS}{\mathcal{S}}

\newcolumntype{L}[1]{>{\raggedright\let\newline\\\arraybackslash\hspace{0pt}}m{#1}}
\newcolumntype{C}[1]{>{\centering\let\newline\\\arraybackslash\hspace{0pt}}m{#1}}
\newcolumntype{R}[1]{>{\raggedleft\let\newline\\\arraybackslash\hspace{0pt}}m{#1}}

\newcommand*\xbar[1]{%
  \hbox{%
    \vbox{%
      \hrule height 0.5pt 
      \kern0.3ex
      \hbox{%
        \kern-0.1em
        \ensuremath{#1}%
        \kern-0.05em
      }%
    }
  }
}

\newtheoremstyle{fullit}
  {\topsep}      
  {\topsep}      
  {\normalfont}  
  {0pt}          
  {\itshape}     
  {.\ }          
  {0pt}          
  {\thmname{#1} \thmnumber{#2}}             

\newtheorem{thm}{Theorem}[section]

\theoremstyle{definition}

\theoremstyle{fullit}
\newtheorem{example}[thm]{Example}

\definecolor{urlssc}{HTML}{161680}
\definecolor{linksc}{HTML}{80164B}
\definecolor{citesc}{HTML}{16804B}
\hypersetup{
  linkcolor  = linksc!80!black,
  citecolor  = citesc!80!black,
  urlcolor   = urlssc,
  colorlinks = true,
}

\tikzset{cross/.style={cross out, draw=black, minimum size=2*(#1-\pgflinewidth), inner sep=0pt, outer sep=0pt},
cross/.default={1pt}}

\DeclareCaptionSubType*[alph]{figure}
\begin{document}
\setlength{\parindent}{0pt}
\vspace*{-1.5cm}

\vspace{1.5cm}
\title{Little String Defects and Bala--Carter Theory}

\vspace{0.35cm}
\author{Nathan Haouzi}
\author{Christian Schmid}

\vspace{0.1cm}
\address[A1,A2]{
Center for Theoretical Physics\\
University of California, Berkeley, USA}
\email[A1]{\href{mailto:nathanhaouzi@berkeley.edu}{nathanhaouzi@berkeley.edu}}
\email[A2]{\href{mailto:cschmid@berkeley.edu}{cschmid@berkeley.edu}}

\vspace{1cm}


\begin{abstract}
We give a physical realization of the Bala--Carter labels that classify nilpotent orbits of semi-simple Lie algebras, for the case $\fg= A, D, E$. We start from type IIB string theory compactified on an $ADE$ singularity and study the six-dimensional (2,0) $\fg$-type little string  on a Riemann surface with punctures. The defects are introduced as D-branes wrapping the 2-cycles of the singularity.  At low energies, the little string becomes the (2,0) conformal field theory of type $\fg$. As an application, we derive the full list of $E_n$ little string defects, and their Bala--Carter label in the CFT limit. Furthermore, we investigate new relations between the quiver gauge theory describing the D-brane defects at low energies, and the weighted Dynkin diagrams of $\fg$. We also give a physical version of the dimension formula of a nilpotent orbit based on its weighted Dynkin diagram.

\end{abstract}
\maketitle
\clearpage

\tableofcontents


\colorlet{brighty}{red!20!yellow!20!white}
\newpage
\section{Introduction}
\label{sec:intro}
String theory predicts the existence of a six-dimensional conformal field theory with $(2,0)$ supersymmetry;  this theory has recently attracted a lot of interest, particularly as a means to study lower dimensional quantum field theories (\cite{Gaiotto:2009we,Gaiotto:2009hg,Witten:2009at}; see also \cite{Chacaltana:2012zy,Balasubramanian:2014dia}). It is labeled by a Lie algebra $\fg$ of $ADE$ type. One can construct it from string theory by taking two limits: Type IIB string theory is first put on an $ADE$ surface $X$. Then, the string coupling $g_s$ is sent to zero; this gives a six-dimensional string theory, called the $(2,0)$ little string. Afterwards, the string mass $m_s$ is taken to infinity (or equivalently, the string length to zero), while keeping the various moduli of the $(2,0)$ string theory fixed. As a result, the theory doesn't have any more scale parameters, and we end up with the $(2,0)$ CFT.\\

However, it proves useful to study the $(2,0)$ little string proper, at finite $m_s$, instead of its CFT limit. Namely, it is  a powerful tool to analyze supersymmetric gauge theories in lower dimensions. For instance, in \cite{Aganagic:2015cta}, the $(2,0)$ little string is studied on a Riemann surface $\cC$, with defects. Specifically, one introduces codimension-two defects as D5 branes that are points on $\cC$ and wrap non-compact 2-cycles of $X$. In the setup of \cite{Haouzi:2016ohr}, which we will use again here for consistency of notations, one further compactifies the little string on a torus $T^2$; the defects are then equivalently described as D3 branes at points on $T^2$, by T-duality.

The $(2,0)$ little string theory with the D3 brane defects turns out to be described by the theory on the branes themselves  (except at the singular origin of the moduli space). At low energies, it is given by a 2d $\cN=(4,4)$  quiver gauge theory; the shape of the quiver is the Dynkin diagram of $\fg$.\\

 In \cite{Haouzi:2016ohr}, we fully classified the defects of the $(2,0)$ little string on $\cC$. The defects are described by a set of weights of $\fg$, subject to some conditions. From this data, one can extract a quiver gauge theory. After taking the string mass $m_s$ to infinity, to get a defect of the $(2,0)$ CFT, one generically loses this low energy gauge theory description of the defects.\footnote{Indeed, the D-brane inverse gauge coupling $\tau$ must go to zero, because $\tau m_s^2$ is a modulus of the $(2,0)$ theory that is kept fixed in the limit.} However, in that limit, one is then able to extract a parabolic subalgebra of $\fg$. This subalgebra characterizes a defect of the $(2,0)$ CFT which agrees with the  known classification in terms of nilpotent orbits \cite{Chacaltana:2012zy,Chacaltana:2010ks,*Chacaltana:2011ze,*Chacaltana:2013oka,*Chacaltana:2014jba,*Chacaltana:2015bna,*Chacaltana:2016shw}. In other words, a given set of D3 branes wrapping 2-cycles of $X$  determines a unique parabolic subalgebra of $\fg$ at low energies. In hindsight, the fact that parabolic subalgebras make an appearance in the CFT limit could have been anticipated. Indeed, in the limit, the brane defects of the little string become the surface defects of $\cN=4$ SYM studied by Gukov and Witten \cite{Gukov:2006jk}. As they explain, these can be described as the sigma model $T^*(G/P)$, with $P$ a parabolic subgroup of the Lie group $G$. A main result of \cite{Haouzi:2016ohr} is that the space $T^*(G/P)$ should be identified with the Coulomb branch of the brane defect theory, when $m_s\rightarrow\infty$.\\

In this paper, we elucidate further the connection between codimension 2 defects of the little string and nilpotent orbits. In the CFT limit, we find that defects realize physically the classification of nilpotent orbits derived by Bala and Carter \cite{bala:1976msaa,*bala:1976msab}. In this note, we will limit our analysis to the case $\fg=A, D, E$, but note that the Bala--Carter classification is in fact applicable to all semi-simple Lie algebras. We derive the Bala--Carter labels for the nilpotent orbits from the weight data that defines the low energy theory on the D3 branes in the little string.

This characterization of defects is recovered in the context of 2d Toda CFT on $\cC$. Indeed,
the AGT correspondence \cite{Alday:2009aq} relates four-dimensional $\cN=2$ theories compactified on a Riemann surface to a two-dimensional $\fg$-Toda conformal field theory on the surface. The Bala--Carter labels can be read off directly from null state conditions at level 1 in the Toda CFT \cite{Kanno:2009ga}.\\

There exists yet another way to classify nilpotent orbits of $\fg$, which turns out to be related to the above defect classification: by the so-called weighted Dynkin diagrams. These arise as a consequence of the Jacobson--Morozov theorem and we will point out an unexpected connection to the quivers arising in the little string theory context. 

Indeed, any nilpotent orbit $\cO$ has a representative $X$ that fits into an $\mathfrak{sl}_2$ triple. This means that there exists a nilpotent $Y$ and a semi-simple $H$ with
\begin{equation}
[H,X]=2X,\qquad [H,Y]=-2Y, \qquad [X,Y]=H.
\end{equation}
By the Jacobson--Morozov theorem, this triple is unique up to conjugation. One then constructs a quiver of Dynkin shape, with the integer entries 0, 1, or 2, from the semi-simple element $H$. In this way, every nilpotent orbit is denoted by a distinct diagram, called a weighted Dynkin diagram. Surprisingly, it turns out that these diagrams can be understood as physical little string defect quivers; the integers 0, 1, or 2 of the weighted Dynkin diagram should then be understood as the rank of unitary flavor symmetry groups. We do not have an interpretation for this observation, but we nonetheless explore some of its consequences.\\

The paper is organized as follows. In section \ref{sec:review}, we review the construction of \cite{Haouzi:2016ohr}, and the classification of surface defects of the $(2,0)$ little string theory; we further recall how one extracts a parabolic subalgebra from a defect, in the CFT limit $m_s\rightarrow \infty$. In section \ref{sec:Bala}, we show that this characterization of defects is precisely what is called the Bala--Carter classification of nilpotent orbit in the mathematics literature \cite{bala:1976msaa,*bala:1976msab}. In section \ref{sec:dynkin}, we review how nilpotent orbits of $\fg$ are also classified by weighted Dynkin diagrams, and we point out a connection to the quivers arising in the little string theory. In appendix \ref{sec:results}, we give the explicit list of defects of $\fg=E_n$ little string and its CFT limit, as an application of the Bala--Carter theory introduced in section \ref{sec:Bala}.

\newpage

\section{Review: Little String Defects and Parabolic Subalgebras}
\label{sec:review}

The $(2,0)$ $ADE$ little string is a six dimensional theory, and therefore has 16 supercharges. Its discovery dates back to \cite{Seiberg:1997zk,Witten:1995zh,Losev:1997hx} (see  \cite{Aharony:1999ks} for a review), and it has been studied more recently in \cite{Kim:2015gha,Bhardwaj:2015oru,DelZotto:2015rca,Lin:2015zea,Hohenegger:2016yuv,Aganagic:2015cta,Hohenegger:2015btj, Aganagic:2016jmx}. It is obtained by considering type IIB string theory on an ADE surface $X$ and sending the string coupling $g_s$ to zero; this decouples the bulk modes of the full type IIB string theory, and keeps only the degrees of freedom supported near $X$. The space $X$ is a hyperk\"ahler manifold, obtained by resolving a ${\mathbb C}^2/\Gamma$ singularity. Here $\Gamma$ is a discrete subgroup of $SU(2)$ that is related to ${\bf g}$ by the McKay correspondence \cite{Reid:1997zy}. 

Since the strings have a finite tension $m_s^2$, the little string theory is not a local QFT. At energies below the string scale $m_s$, it reduces to a $(2,0)$ 6d CFT. 

\subsection{Brane Defects of the Little String}

We compactify the $(2,0)$ little string theory on a Riemann surface $\cC$, which we take to be the complex plane. Then, $X\times \cC$ is a solution of the full type IIB string theory. Codimension-two defects are introduced as branes that are points on $\cC$ and fill the four remaining directions $\mathbb{C}^2$. These are D5 branes in IIB string theory, wrapping non-compact 2-cycles in $X$ and  $\mathbb{C}^2$ \cite{Aganagic:2015cta}. Their tension remains finite in the little string limit. Equivalently, after further compactifying the theory on $T^2$ and using T-duality, the defects are described by D3 branes at points on $T^2$. This viewpoint has the advantage of giving the little string theory origin of the surface defects studied by Gukov and Witten \cite{Gukov:2006jk}. Indeed, after $T^2$ compactification and at energies far below $m_s$, the little string theory becomes 4d $\cN=4$ SYM \cite{Vafa:1997mh}, and the D3 branes become codimension two defects. In particular, S-duality of $\cN=4$ SYM with surface defects originates from $T^2$-duality of the little string. For  details, see \cite{Haouzi:2016ohr}.

The dynamics of the $(2,0)$ little string theory on $\cC\times  \mathbb{C}\times T^2$, with an arbitrary collection of D3 brane defects at points on $\cC\times T^2$, is captured by the theory on the branes themselves.\\

Because  $\cC$ has a flat metric, the theory on the D3 branes  at low energies is a two-dimensional quiver gauge theory, of shape the Dynkin diagram of $\fg$ \cite{Douglas:1996sw}. The theory has 2d $\cN=(4,4)$ supersymmetry, since the D3 branes break half the supersymmetry. As in \cite{Aganagic:2015cta, Haouzi:2016ohr}, we are interested in the class of D3 branes that retain some conformal invariance in the resulting low energy theory (the terminology here is borrowed from four dimensions, before $T^2$ compactification). This corresponds to a very specific choice of non-compact 2 cycles of $X$ wrapped by the D3 branes. The moduli of the 6d $(2,0)$ theory become gauge couplings of the 2d theory. The positions of the D3 branes on $\cC$ wrapping  non-compact two-cycles  of $X$ become the masses of  fundamental hypermultiplets. Finally, the positions of the D3 branes on $\cC$ wrapping  compact two-cycles  of $X$ become Coulomb moduli.

In the rest of this paper, we study the quiver gauge theory that describes this low energy limit of D3 branes  that wrap  2-cycles of the ALE surface $X$ times $\mathbb{C}$ -- we will call this theory $T^{2d}$.
\\

To specify the charge of D3 branes wrapping non-compact two-cycles of $X$, we pick a class $[S^*]$ in the relative homology $H_2(X, \partial X; {\mathbb Z}) = \Lambda_*$; by the McKay correspondence, this homology group is identified with the (co-)weight lattice of $\bf g$:
\beq\label{ncomp}
[S^*] = -\sum_{a=1}^{n} \, m_a \, w_a \;\;  \in \Lambda_*\; ,
\eeq
with  $w_a$ the fundamental weights,\footnote{These are the weights that are represented by unit vectors in the Dynkin basis.} $m_a$ non-negative integers, and $n=\mbox{rank}(\fg)$.  As explained, we want to preserve conformal invariance in the low energy quiver theory (in a 4d sense). This is equivalent to a vanishing D3 brane flux at infinity.  We therefore need to add additional D3 branes wrapping a compact homology class $[S]$ in $H_2(X, {\mathbb Z})=\Lambda$. This homology group is identified with the root lattice of $\bf g$:
\beq\label{comp}
[S] = \sum_{a=1}^n  \,d_a\,\alpha_a\;\;  \in  \Lambda.
\eeq
Here, $\alpha_a$ are the simple positive roots of $\fg$, and $d_a$ are non-negative integers
such that
\beq\label{conf}
[S+S^*] =0.
\eeq
If $S+S^*$ vanishes in homology, then $\# (S_a \cap (S+S_*))$ vanishes as well, for all $a$.  We therefore rewrite \eqref{conf} as
\beq\label{conformal}
\sum_{b=1}^n C_{ab} \;d_b = m_a\; ,
\eeq
with $C_{ab}$ the Cartan matrix of $\fg$.

Going to the Higgs branch of the low energy quiver theory, the gauge group $\prod_{a=1}^n U(d_a)$ breaks to $\prod_{a=1}^n U(1)$. There, all the D3 branes recombine and form a configuration of D3 branes wrapping a set of non-compact cycles $S_i^*$; their homology classes $\omega_i$ live in the weight lattice $\Lambda^* = H_2(X, \partial X; {\mathbb Z})$:
\beq\label{weightsfr}
\omega_i  = [S_i^*] \qquad \in\Lambda^*.
\eeq
In what follows, the weights $\omega_i$ will all belong to fundamental representations of $\fg$. Different configurations of the weights $\omega_i$ end up classifying the defects of the little string \cite{Haouzi:2016ohr}. Every one of these $\omega_i$'s comes from one of the non-compact D3 branes on $S^*$.
For the branes to bind, the position of each compact brane has to coincide with the position of one of the non-compact D3 branes on $\cC$. As we mentioned, the positions of non-compact D3 branes are mass parameters of the quiver gauge theory, while the positions of compact D3 branes on $\cC$ are Coulomb moduli. Whenever a compact brane and a non-compact brane meet on $\cC$, the corresponding fundamental hypermultiplet becomes massless and can get a vacuum expectation value. These vevs describe the root of the Higgs branch.\\

Thus, on the root of the Higgs branch, we can write the weights $\omega_i$ in the following form:

$$\omega_i=-w_a+\sum_{b=1}^n k_{ib}\, \alpha_b,$$
where $-w_a$ is a negative fundamental weight, the $k_{ib}$ are non-negative integers, and the $\alpha_b$ are positive simple roots (from bound compact D3 branes). The collection of weights
\begin{equation}
\label{WS}
{\cW}_{\cS} = \{ \omega_i\}
\end{equation}
we get must be such that it accounts for all the D3 brane charges in $[S^*]$ and in $[S]$. One simple consequence is that the number of $\omega_i$'s is the total rank of the 2d flavor group, $\sum_{a=1}^n m_a$. Because the net D3 brane charge is zero, $[S+S^*]=0$, so that

\[
\sum_{\omega_i\in{\cW}_{\cS}}\omega_i=0,
\]
which is equivalent to \eqref{conformal}. In this note, we take the set ${\cW}_{\cS}$ to have size at  most $n+1$, as a $n+1$ weights are sufficient to describe the most generic defect of the $\fg$-type little string. We emphasize that all weights of ${\cW}_{\cS}$ we consider will be taken in fundamental representations.\\

As an example, the full puncture of $\fg=D_4$ is featured below.\\

\begin{example}
\label{ex:d4ex}
Let $\cW_S$ be the following set of weights of $D_4$:
\begin{alignat*}{2}
\omega_1&=[\phantom{-}1,\phantom{-}0,\phantom{-}0,\phantom{-}0]& &=-w_1+2\alpha_1+2\alpha_2+\alpha_3+\alpha_4,\\
\omega_2&=[-1,\phantom{-}1,\phantom{-}0,\phantom{-}0]& &=-w_1+\alpha_1+2\alpha_2+\alpha_3+\alpha_4,\\
\omega_3&=[\phantom{-}0,-1,\phantom{-}1,\phantom{-}1]& &=-w_1+\alpha_1+\alpha_2+\alpha_3+\alpha_4,\\
\omega_4&=[\phantom{-}0,\phantom{-}0,-1,\phantom{-}0]& &=-w_3,\\
\omega_5&=[\phantom{-}0,\phantom{-}0,\phantom{-}0,-1]& &=-w_4.
\end{alignat*}
Note that these weights add up to zero, as they should.
Written as above, these weights define a 2d quiver gauge theory, shown in Figure \ref{fig:d4full}.
\begin{figure}[htbp]
\begin{center}
\begin{tikzpicture}
\begin{scope}[auto, every node/.style={minimum size=0.75cm}]
\def \spac {1cm}

\node[circle, draw](k1) at (0,0) {$4$};
\node[circle, draw](k2) at (1*\spac,0) {$5$};
\node[circle, draw](k3) at (2*\spac,0.6*\spac) {$3$};
\node[circle, draw](k4) at (2*\spac,-0.6*\spac) {$3$};

\node[draw, inner sep=0.1cm,minimum size=0.67cm](N1) at (0*\spac,\spac) {$3$};
\node[draw, inner sep=0.1cm,minimum size=0.67cm](N3) at (3*\spac,0.6*\spac) {$1$};
\node[draw, inner sep=0.1cm,minimum size=0.67cm](N4) at (3*\spac,-0.6*\spac) {$1$};

\draw[-] (k1) to (k2);
\draw[-] (k2) to (k3);
\draw[-] (k2) to (k4);

\draw (k1) -- (N1);
\draw (k3) -- (N3);
\draw (k4) -- (N4);

\end{scope}
\end{tikzpicture}
\end{center}
\caption{The quiver describing a full puncture for $\fg=D_4$. Unitary gauge groups are circles, while unitary flavor groups are squares.}
\label{fig:d4full}
\end{figure}
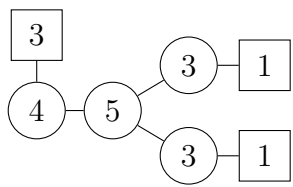
\end{example}

This quiver gauge theory description is only valid at finite  $m_s$. 
After taking the $m_s\to\infty$ limit, the $(2,0)$ little string becomes the $(2,0)$ CFT of type $\fg$; in general, this results in the loss of the Lagrangian description of the defect. In particular, the Coulomb branch dimension, previously equal to $\sum_{a=1}^n d_a$, generically decreases for $\fg=D_n, E_n$. This loss of Coulomb moduli is not unexpected, since in the limit, the theory loses degrees of freedom. We denote the resulting theory by $T^{2d}_{m_s\rightarrow \infty}$. 

In this setup, it turns out there is a remarkably simple classification of defects, divided into two categories, both for the little string defect $T^{2d}$ and for its $(2,0)$ CFT limit $T^{2d}_{m_s\rightarrow \infty}$. We now review it.\\

\subsection{Polarized Defects}

Pick a weight $\omega$ in the $a$-th fundamental representation of $\fg$, labeled by the fundamental weight $w_a$. Suppose that $\omega$ is in the Weyl group orbit of $-w_a$; then if all weights of  $\cW_{\cS}$  satisfy this condition,  we call the resulting 2d theory $T^{2d}$ on the D3 branes  \emph{polarized}.
In the CFT limit, such a theory $T^{2d}_{m_s\rightarrow\infty}$ distinguishes a parabolic subalgebra.\\

First remember that a \emph{Borel subalgebra} of $\fg$ is a maximal solvable subalgebra of $\fg$. It can always be written as $\fb=\fh\oplus\fm$, with $\fh$ a Cartan subalgebra of $\fg$ and $\fm=\sum_{\alpha\in\Phi^+}\fg_\alpha$, where $\fg_\alpha$ are the root spaces associated to a set of positive roots $\Phi^+$. In what follows, we fix this set of positive roots, or equivalently, the Borel subalgebra $\fb$, for each $\fg$.

A \emph{parabolic subalgebra} $\fp_\Theta$ is a subalgebra of $\fg$ that can be written as $$\fp_\Theta=\fl_\Theta\oplus\fn_\Theta.$$ Here, $\Theta$ is a subset of the set of simple positive roots of $\fg$. Here, $\fl_\Theta=\fh\ \oplus\sum_{\alpha\in\langle\Theta\rangle}\fg_\alpha$ is called a Levi subalgebra, and  $\fn_\Theta=\sum_{\alpha\in\Phi^+ \backslash \langle\Theta\rangle^+}\fg_\alpha$, is called the nilradical of $\fp_\Theta$. $\langle\Theta\rangle$ is the subroot system of $\fg$ generated by $\Theta$, and $\langle\Theta\rangle^+$ is made of the positive roots of $\langle\Theta\rangle$.

Note that this implies $\fn_{\Theta}\cong \fg/\fp_{\Theta}$. See below for an example:\\

\begin{example}
Consider $\fg=A_2$ in the first fundamental representation, which is the defining representation of $\mathfrak{sl}_3$. The root space $\fg_\alpha$ associated to a root $\alpha=h_i-h_j$ is $\mathbb{C}E_{ij}$, where $E_{ij}$ is a matrix that has a $1$ in the $i$-th row and $j$-th column, and zeros everywhere else. (See Table \ref{tab:a2ex} below for an illustration.)

\tcbset{enhanced,size=fbox,nobeforeafter,tcbox raise=-2.3mm,colback=white,colframe=white}
\begin{table}[htp]
\renewcommand{\arraystretch}{1.2}
\begin{center}
\begin{tabular}{c|c|c|c}
$\Theta$&$\fp_\Theta$&$\fl_\Theta$&$\fn_\Theta$\\
\hline
$\varnothing$&$\begin{pmatrix}
*&\tcbox[colframe=lime]{*}&\tcbox[borderline={0.2mm}{0mm}{green!95!black,dashed}]{*}\\
0&*&\tcbox[colframe=brown]{*}\\
0&0&*
\end{pmatrix}$
&
$\begin{pmatrix}
*&0&0\\
0&*&0\\
0&0&*
\end{pmatrix}$&$\begin{pmatrix}
0&*&*\\
0&0&*\\
0&0&0
\end{pmatrix}$
\end{tabular}\\
\end{center}
\begin{flushleft}
\hspace*{9em}\tcbox[colframe=lime]{\phantom{*}}: $\alpha_1$\hspace{3em} \tcbox[borderline={0.2mm}{0mm}{green!95!black,dashed}]{\phantom{*}}: $(\alpha_1+\alpha_2)$\\
\hspace*{9em}\tcbox[colframe=brown]{\phantom{*}}: $\alpha_2$
\end{flushleft}
	\caption{The colored boxes label the root spaces associated to the positive roots of $\fg$, as indicated.}
\label{tab:a2ex}
\end{table}

The Borel subalgebra we choose is
$$
\fb=\begin{pmatrix}
*&*&*\\
0&*&*\\
0&0&*
\end{pmatrix}.
$$

Then the Cartan subalgebra has the form
\begin{equation}
\fh=\begin{pmatrix}
*&0&0\\
0&*&0\\
0&0&*
\end{pmatrix}.
\end{equation}

The full list of all $A_2$ parabolic subalgebras is:
\begin{align*}
\fp_\varnothing=\fb&=\begin{pmatrix}
*&*&*\\
0&*&*\\
0&0&*
\end{pmatrix},\,
\fp_{\{\alpha_1\}}=\begin{pmatrix}
*&*&*\\
*&*&*\\
0&0&*
\end{pmatrix},\,
\fp_{\{\alpha_2\}}=\begin{pmatrix}
*&*&*\\
0&*&*\\
0&*&*
\end{pmatrix},\,
\fp_{\{\alpha_1,\alpha_2\}}=\begin{pmatrix}
*&*&*\\
*&*&*\\
*&*&*
\end{pmatrix},
\end{align*}

with Levi decomposition:
\begin{align}
\fp_\varnothing&=\begin{pmatrix}
*&*&*\\
0&*&*\\
0&0&*
\end{pmatrix}=\begin{pmatrix}
*&0&0\\
0&*&0\\
0&0&*
\end{pmatrix}\oplus
\begin{pmatrix}
0&*&*\\
0&0&*\\
0&0&0
\end{pmatrix}=\fl_\varnothing\oplus\fn_\varnothing,\\
\fp_{\{\alpha_1\}}&=\begin{pmatrix}
*&*&*\\
*&*&*\\
0&0&*
\end{pmatrix}=\begin{pmatrix}
*&*&0\\
*&*&0\\
0&0&*
\end{pmatrix}\oplus
\begin{pmatrix}
0&0&*\\
0&0&*\\
0&0&0
\end{pmatrix}=\fl_{\{\alpha_1\}}\oplus\fn_{\{\alpha_1\}},\\
\fp_{\{\alpha_2\}}&=\begin{pmatrix}
*&*&*\\
0&*&*\\
0&*&*
\end{pmatrix}=\begin{pmatrix}
*&0&0\\
0&*&*\\
0&*&*
\end{pmatrix}\oplus
\begin{pmatrix}
0&*&*\\
0&0&0\\
0&0&0
\end{pmatrix}=\fl_{\{\alpha_2\}}\oplus\fn_{\{\alpha_2\}},\\
\fp_{\{\alpha_1,\alpha_2\}}&=\begin{pmatrix}
*&*&*\\
*&*&*\\
*&*&*
\end{pmatrix}=\begin{pmatrix}
*&*&*\\
*&*&*\\
*&*&*
\end{pmatrix}\oplus
\begin{pmatrix}
0&0&0\\
0&0&0\\
0&0&0
\end{pmatrix}=\fl_{\{\alpha_1,\alpha_2\}}\oplus\fn_{\{\alpha_1,\alpha_2\}}.
\end{align}

\end{example}

Parabolic subalgebras of $\fg$ are recovered from  D3 brane defects in the following way \cite{Haouzi:2016ohr}:
Given a  set of weights
\begin{equation*}
{\cW}_{\cS} = \{ \omega_i\},
\end{equation*}
we want to construct a set $\Theta$ as the subset of all simple roots that have a vanishing inner products with all the weights $\omega_i$. To achieve this, we have to choose a set in the Weyl group orbit of $\cW_{\cS}$ for which $|\Theta|$ is maximal.\footnote{The Weyl group acts on all weights of $\cW_{\cS}$ simultaneously.} Such a set of weights is called \emph{distinguished}.\\

Furthermore, the nilradical $\fn_{\Theta}=\fg/\fp_{\Theta}$ specifies the Coulomb branch of  $T^{2d}_{m_s\rightarrow \infty}$. 
It is given by the direct sum of the root spaces associated to the positive roots $e_{\gamma}$ of $\fg$ for which:   The positive roots $e_\gamma$ that satisfy
\beq
\langle e_{\gamma}, \omega_i\rangle <0\text{\footnotemark}
\eeq
\footnotetext{equivalently, $\langle e_{\gamma}, \omega_i\rangle >0$.}
for at least one weight $\omega_i\in\cW_{\cS}$. The inner product $\langle \cdot, \cdot \rangle$ is the Killing form of $\fg$.
The dimension of the Coulomb branch of $T^{2d}_{m_s\rightarrow \infty}$ can therefore be conveniently recovered from this prescription. In the little string context, at finite $m_s$, the dimension of the Coulomb branch of $T^{2d}$ is instead given by
$$\sum\limits_{\langle e_\gamma,\omega_i\rangle<0} \left\vert\langle e_\gamma,\omega_i\rangle\right\vert,
$$
where the sum runs over all positive roots $e_\gamma$ and all weights $\omega_i$ of $\cW_{\cS}$ satisfying $\langle e_\gamma,\omega_i\rangle<0$.\\

\begin{example}[$D_4$ example]
	We will calculate the Coulomb branch dimension of the full puncture theory of example \ref{ex:d4ex}, both for $T^{2d}$ and $T^{2d}_{m_s\rightarrow \infty}$.\\
The positive roots $\Phi^+$ of $D_4$ are
\begin{equation*}
\begin{split}
(&h_1+h_2,\;h_1+h_3,\;h_2+h_3,\;h_1+h_4,\;h_1-h_4,\;h_2+h_4,\\
&h_2-h_4,\;h_1-h_3,\;h_2-h_3,\;h_3+h_4,\;h_3-h_4,\;h_1-h_2)
\end{split}
\end{equation*}
The negative inner products of each of these roots with the weights in $\cW_\cS$ are given in the following table, where all positive inner products were replaced by $0$:
\begin{alignat*}{2}
\langle \Phi^+,\omega_1\rangle&=(\phantom{-}0,\phantom{-}0,\phantom{-}0,\phantom{-}0,\phantom{-}0,\phantom{-}0,\phantom{-}0,\phantom{-}0,\phantom{-}0,\phantom{-}0,\phantom{-}0,\phantom{-}0),\\
\langle \Phi^+,\omega_2\rangle&=(\phantom{-}0,\phantom{-}0,\phantom{-}0,\phantom{-}0,\phantom{-}0,\phantom{-}0,\phantom{-}0,\phantom{-}0,\phantom{-}0,\phantom{-}0,\phantom{-}0,-1),\\
\langle \Phi^+,\omega_3\rangle&=(\phantom{-}0,\phantom{-}0,\phantom{-}0,\phantom{-}0,\phantom{-}0,\phantom{-}0,\phantom{-}0,-1,-1,\phantom{-}0,\phantom{-}0,\phantom{-}0),\\
\langle \Phi^+,\omega_4\rangle&=(-1,-1,-1,\phantom{-}0,-1,\phantom{-}0,-1,\phantom{-}0,\phantom{-}0,\phantom{-}0,-1,\phantom{-}0),\\
\langle \Phi^+,\omega_5\rangle&=(-1,-1,-1,-1,\phantom{-}0,-1,\phantom{-}0,\phantom{-}0,\phantom{-}0,-1,\phantom{-}0,\phantom{-}0).
\end{alignat*}
Adding the absolute value of all these entries gives 15, the dimension of the Coulomb branch of $T^{2d}$. Comparing this to the quiver in Figure \ref{fig:d4full}, this is indeed correct.

Furthermore, we can see that all 12 positive roots have a negative inner product with at least one of the weights. Thus, the Coulomb branch of $T^{2d}_{m_s\to\infty}$ has (complex) dimension 12.

The set $\cW_\cS$ is distinguished, and one can see immediately that $\Theta=\varnothing$. So the parabolic subalgebra associated to this defect is all of $D_4$.\\
\end{example}

The parabolic subalgebra $\fp_{\Theta}$ also makes an appearance in the  context of the AGT correspondence \cite{Alday:2009aq}. This duality predicts that codimension-two defects of the $\fg$-type 6d $(2,0)$ CFT on $\cC$ should be classified from the point of view of a 2d conformal field theory on $\cC$, called $\fg$-Toda conformal field theory. This CFT has an extended conformal symmetry, called a ${\cW}({{\fg}})$-algebra symmetry. Defects of the 6d $(2,0)$ conformal field theory correspond to certain vertex operator insertions at points on $\cC$ in the 2d Toda CFT.\\

In fact, the elements of the Cartan subalgebra $\fh\subset \fg$  define the highest weight states $|\vec\beta\rangle$ of the  $\cW(\fg)$-algebra. In terms of the weights $\omega_i$ of the algebra $\fg$, we write $\vec\beta=\sum_{i=1}^{|\cW_{\cS}|}\beta_i\; \omega_i$.
Polarized defects  then turn out to be characterized in Toda theory by level 1 null states \cite{Kanno:2009ga}. Explicitly, a highest weight state $|\vec\beta\rangle$ generating a degenerate representation of $\cW(\fg)$ satisfies
\[
\vec\beta\cdot\vec\alpha_i=0 \quad \forall \vec\alpha_i\in\Theta
\]
for a subset of simple roots $\Theta$. Then the parabolic subalgebra associated to $\vec{\beta}$ is just $\fp_\Theta$, which defines a theory $T^{2d}_{m_s\rightarrow \infty}$. In particular, note that a set of positive simple roots characterizes a level 1 null state of Toda.\\

\subsection{Unpolarized Defects}

All the fundamental representations of $A_n$ are minuscule, so all $A_n$ codimension-two defects are polarized. However, in the $D_n$ and $E_n$ cases, it can also happen that some weight of $\cW_{\cS}$ fails to satistfy the conditions to produce a polarized defect. Namely, two things can happen: First, $\cW_{\cS}$ can turn out to be the set containing the zero weight only (possibly multiple times, as will occur first for $\fg=E_7$, see Appendix \ref{sec:zero}). Second, $\cW_{\cS}$ can contain a nonzero weight $\omega$ in the representation generated by (minus) a fundamental weight $-w_a$ without being in the Weyl orbit of $-w_a$. Either way, additional data is needed beyond simply specifying the set $\cW_{\cS}$. Therefore, weights that make up an unpolarized defect need to be given a subscript denoting (minus) the representation they are taken in \footnote{the "minus" here is because every weight we consider is written as $\omega=-w_a+\ldots$}. In the CFT limit $m_s\rightarrow \infty$, no parabolic subalgebra of $\fg$ is singled out. Why this is so will be explained in the next section.\\

\FloatBarrier
\begin{figure}[htpb]
	\begin{center}
\begin{tikzpicture}[font=\footnotesize]
\draw[dashed, line width=1.5pt, green!70!black] (0,3.5) -- (0,-0.5);
\draw (-2.8,3) -- (-2.8,0);
\draw (-2,3) -- (-2,0);
\draw (-1.2,3) -- (-1.2,0);
\draw (-0.4,3) -- (-0.4,0);
\draw (0.4,3) -- (0.4,0);
\draw (1.2,3) -- (1.2,0);
\draw (2,3) -- (2,0);
\draw (2.8,3) -- (2.8,0);

\node(c2) at (1.6,1.5) {};
\node at ($(c2)+(0,0.5)$) {\color{red}$\omega_1$};

\draw[-,line width=5pt, red!70!yellow] ($(c2)+(-0.2,0.2)$) -- ($(c2)+(0.2,-0.2)$);
\draw[-,line width=5pt, red!70!yellow] ($(c2)+(-0.2,-0.2)$) -- ($(c2)+(0.2,0.2)$);

\node(c4) at (-1.6,1.5) {};
\node at ($(c4)+(0,0.5)$) {\color{red}$\omega_1$};

\draw[-,line width=5pt, red!70!yellow] ($(c4)+(-0.2,0.2)$) -- ($(c4)+(0.2,-0.2)$);
\draw[-,line width=5pt, red!70!yellow] ($(c4)+(-0.2,-0.2)$) -- ($(c4)+(0.2,0.2)$);

\draw[red!70!yellow] (-2.8,1.6) -- (2,1.6);
\draw[red!70!yellow] (-2,1.4) -- (2.8,1.4);

\node[align=left,text width=6.8cm] at (7,1.5) {$\omega_1: [0,0,0,0]=\color{red!70!yellow}-w_2+\alpha_1+2\alpha_2+\alpha_3+\alpha_4$};

\end{tikzpicture}
	\end{center}
	\caption{The zero weight $[0,0,0,0]$ of the $D_4$ algebra is the simplest example of how one  constructs an unpolarized defect of the little string; on the left is pictured the type IIB brane engineering of the weight. NS5 branes are vertical black lines, D5 branes are red crosses, and D3 branes are horizontal red lines. The green dotted line produces a $\mathbb{Z}_2$-orbifold of an $A_7$ theory, realizing the $D_4$ theory. The resulting defect will be unpolarized because  $[0,0,0,0]$ belongs in the $[0,1,0,0]$ representation, but is not in the Weyl group orbit of that weight.}
	\label{fig:D4unpolarized}
\end{figure}
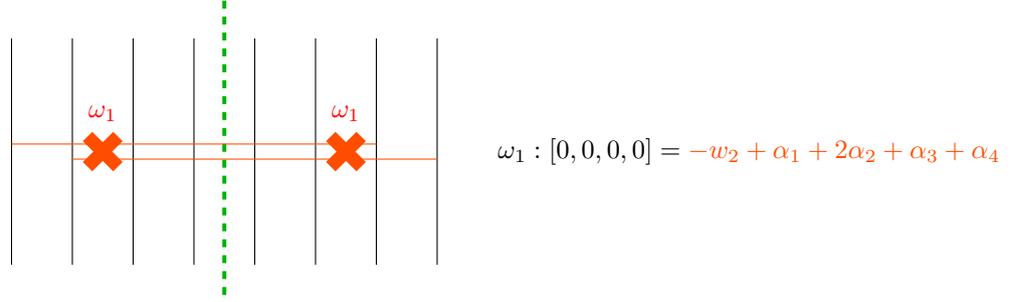
\begin{figure}[hbpt]
	\begin{center}
		\begin{tikzpicture}[font=\footnotesize]
\draw[dashed, line width=1.5pt, green!70!black] (0,3.5) -- (0,-3.5);
\draw (-3.6,3) -- (-3.6,-3);
\draw (-2.8,3) -- (-2.8,-3);
\draw (-2,3) -- (-2,-3);
\draw (-1.2,3) -- (-1.2,-3);
\draw (-0.4,3) -- (-0.4,-3);
\draw (0.4,3) -- (0.4,-3);
\draw (1.2,3) -- (1.2,-3);
\draw (2,3) -- (2,-3);
\draw (2.8,3) -- (2.8,-3);
\draw (3.6,3) -- (3.6,-3);

\draw[white, line width=10pt] (-3.7,0) -- (3.7,0);

\node(c1) at (1.6,-1.5) {};
\node at ($(c1)+(0,0.5)$) {\color{red!70!yellow}$\omega_1$};

\draw[-,line width=5pt, red!70!yellow] ($(c1)+(-0.2,0.2)$) -- ($(c1)+(0.2,-0.2)$);
\draw[-,line width=5pt,red!70!yellow] ($(c1)+(-0.2,-0.2)$) -- ($(c1)+(0.2,0.2)$);

\node(c3) at (-1.6,-1.5) {};
\node at ($(c3)+(0,0.5)$) {\color{red!70!yellow}$\omega_1$};

\draw[-,line width=5pt, red!70!yellow] ($(c3)+(-0.2,0.2)$) -- ($(c3)+(0.2,-0.2)$);
\draw[-,line width=5pt, red!70!yellow] ($(c3)+(-0.2,-0.2)$) -- ($(c3)+(0.2,0.2)$);

\node(c2) at (3.2,2) {};
\node at ($(c2)+(0,0.5)$) {\color{red}$\omega_1$};

\draw[-,line width=5pt, red] ($(c2)+(-0.2,0.2)$) -- ($(c2)+(0.2,-0.2)$);
\draw[-,line width=5pt, red] ($(c2)+(-0.2,-0.2)$) -- ($(c2)+(0.2,0.2)$);

\node(c4) at (-3.2,2) {};
\node at ($(c4)+(0,0.5)$) {\color{red}$\omega_1$};

\draw[-,line width=5pt, red] ($(c4)+(-0.2,0.2)$) -- ($(c4)+(0.2,-0.2)$);
\draw[-,line width=5pt, red] ($(c4)+(-0.2,-0.2)$) -- ($(c4)+(0.2,0.2)$);

\draw[red!70!yellow] (-2.8,-1.4) -- (2,-1.4);
\draw[red!70!yellow] (-2,-1.6) -- (2.8,-1.6);

\node[align=left,text width=6.8cm] at (7.5,2) {$\omega_1: [-1,0,0,0,0]=\color{red}-w_1$};
\node[align=left,text width=6.8cm] at (7.5,-1.5) {$\omega_1: [-1,0,0,0,0]_3=\color{red!70!yellow}-w_3+\alpha_2+2\alpha_3+\alpha_4+\alpha_5$};

\end{tikzpicture}
	\end{center}
	\caption{The weight $[-1,0,0,0,0]$ of $D_5$, with the corresponding type IIB brane engineering on the left. $[-1,0,0,0,0]$  can be written in two ways. First, by placing a D5 brane between the  two leftmost NS5 branes (top), the weight is written appropriately to characterize a polarized defect. This is so because $[-1,0,0,0,0]$ not only belongs in the $[1,0,0,0,0]$ representation, it is also in the Weyl group orbit of that weight. By placing the D5 brane between a different set of NS5 branes (bottom), we will obtain instead an unpolarized defect. This is so because $[-1,0,0,0,0]$ belongs in the $[0,0,1,0,0]$ representation, but is not in the Weyl group orbit of that weight. An additional subscript is added to the weight in this case, denoting (minus) the representation it belongs in.}
	\label{fig:D5unpolarized}
\end{figure}
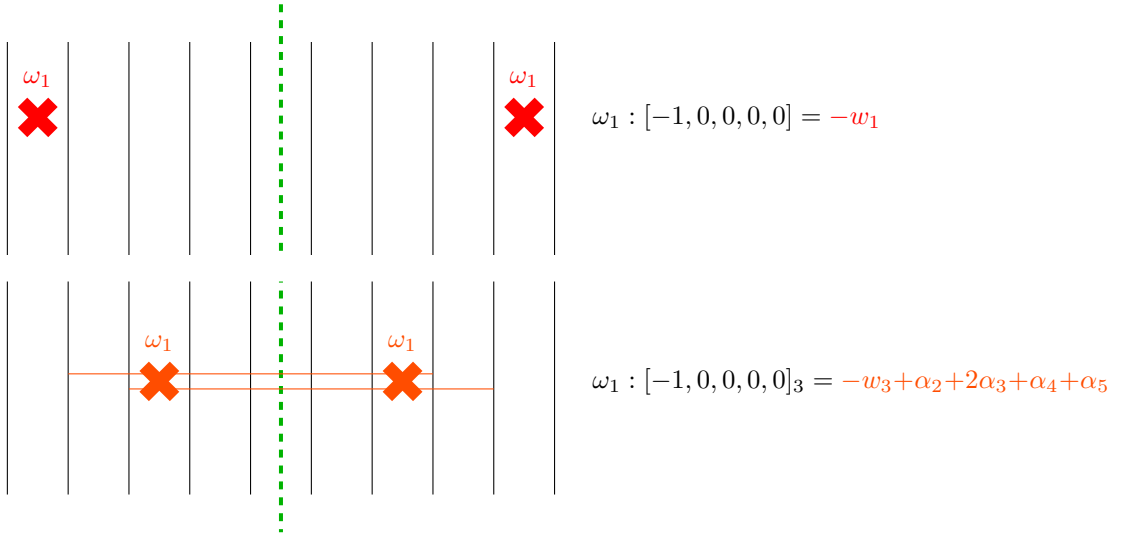
\FloatBarrier

Finally, let us note that the labeling of defects we presented also has implications in the context of the triality worked out in \cite{Aganagic:2013tta, Aganagic:2014oia, Aganagic:2015cta}: the partition function of the $(2,0)$ $\fg=ADE$ little string on $\cC$ with polarized brane defects is equal to a $q$-deformation of the $\fg$-Toda CFT conformal block on $\cC$, with vertex operators determined by positions and the weights. For unpolarized brane defects, the relation to $q$-deformed Toda fails, and other methods have to be used to recover the duality \cite{Haouzi:2016ohr}.\\

Having reviewed the classification of little string defects and their CFT limit, we now proceed to show that their characterization, along with their organization into two classes, polarized and unpolarized, is precisely the Bala--Carter labeling of nilpotent orbits that appears in mathematics \cite{bala:1976msaa,*bala:1976msab}.\\

\section{Bala--Carter Classification}
\label{sec:Bala}

Defects of the 6d $(2,0)$ $\fg$-type CFT have been studied in the literature \cite{Chacaltana:2012zy} in terms of \emph{nilpotent orbits} of the algebra: an element $X\in \fg$ is called nilpotent if the matrix representing $X$ is nilpotent.\footnote{We assume that a fixed faithful representation is chosen for $\fg$.} If $X$ is nilpotent, every element in the $G$-adjoint orbit $\cO_X$ is nilpotent -- this is called a nilpotent orbit. Nilpotent orbits are directly related to the parabolic subalgebras we have been considering. Indeed, given a parabolic subalgebra with Levi decomposition $\fp=\fl\oplus\fn$, the nilpotent orbit $\mathcal{O}_\fl$ associated to $\fp$ is the maximal orbit containing a representative $X\in \mathcal{O}_\fl$ for which $X\in \fn$.

Many of the interesting properties of nilpotent orbits are related to the existence of a duality map: The \emph{Spaltenstein map} \cite{Spaltenstein:1982} sends the set of nilpotent orbits of a simply-laced Lie algebra $\fg$ to itself, and reorganizes them.

For $\fg=A_n$, nilpotent orbits are in one-to-one correspondence with integer partitions of $n+1$ (or Young diagrams); for $\fg=D_n$, they can also be labeled by Young diagrams with $2n$ boxes that satisfy certain conditions (cf.\ the textbook \cite{Collingwood:1993} for more details.) However no such classification in terms of Young diagrams exists for the $E_n$ algebras.\\

It proves fruitful instead to ignore  Young diagrams altogether and resort to the classification of Bala and Carter \cite{bala:1976msaa,*bala:1976msab}, which is valid for any semi-simple Lie algebra. We will see next that this is the natural language to describe the D3 brane defects in the low energy limit.

\subsection{Bala--Carter Labeling  of Nilpotent Orbits}
\label{ssec:balalabel}

Since there are only finitely many orbits in $\fg$, we want to find a convenient way of classifying them. One such classification scheme uses \emph{Levi subalgebras} of $\fg$:

Recall that a Levi subalgebra $\fl$ of $\fg$ is a subalgebra of $\fg$ that can be written as:

$$
\fl=\fh\oplus\bigoplus\limits_{\alpha\in\langle\Theta\rangle}\fg_\alpha,
$$

where $\fh$ is a Cartan subalgebra of $\fg$, $\Theta$ is an arbitrary subset of the simple roots of $\fg$, and $\fg_\alpha$ is the root space associated to a root in the additive closure of $\Theta$.

Then, the idea of the Bala--Carter \cite{bala:1976msaa,*bala:1976msab} classification of nilpotent orbits is to label an orbit $\mathcal{O}$ by the smallest Levi subalgebra that contains a representative of $\mathcal{O}$. This is always unique if $\fg=A_n$, but for other algebras, two different nilpotent orbits can be associated to the same minimal Levi subalgebra.

In general, the following result holds:
A nilpotent orbit $\mathcal{O}$ is uniquely specified by a Levi subalgebra $\fl\subset\fg$ and a certain (distinguished) parabolic subalgebra of $[\fl,\fl]$. These two algebras give the \emph{Bala--Carter label} of $\cO$. A parabolic subalgebra $\fp=\fl^\prime\oplus\fu$, with nilradical $\fu$ and Levi part $\fl^\prime$ is distinguished if $\dim \fl^\prime=\dim \left(\fu/[\fu,\fu]\right)$. One such distinguished parabolic subalgebra is the Borel subalgebra of $\fl$. The nilpotent orbit associated to it is called the \emph{principal nilpotent orbit} of $\fl$. 

Whenever the minimal Levi subalgebra associated to $\cO$ only contains one distinguished parabolic subalgebra (so when $\fl$ uniquely specifies $\cO$) we call the orbit $\cO$ polarized. For simplicity of notation, the Bala--Carter label for such an orbit is just $\fl$. For an unpolarized orbit, it is given by $\fl$ and an additional label \footnote{The additional label specifies the number of simple roots that live in a Levi subalgebra of $\fp$.}.

\begin{example}
For $\fg=A_3$, consider the orbit of the element
$$X=\begin{pmatrix}
0&1&0&0\\
0&0&0&0\\
0&0&0&1\\
0&0&0&0
\end{pmatrix}.
$$
The algebra $\mathfrak{sl}_4$ has five different (conjugacy classes of) Levi subalgebras, corresponding to the five integer partitions of 4. $X$ itself obviously is an element of the Levi subalgebra $\fl_{\{\alpha_1,\alpha_3\}}:$

$$\fl_{\{\alpha_1,\alpha_3\}}=\begin{pmatrix}
*&*&0&0\\
*&*&0&0\\
0&0&*&*\\
0&0&*&*
\end{pmatrix}.
$$
This algebra contains
$$\fl_{\{\alpha_1\}}=\begin{pmatrix}
*&*&0&0\\
*&*&0&0\\
0&0&*&0\\
0&0&0&*
\end{pmatrix}.
$$
Since every element in any conjugacy class of $\fl_{\{\alpha_1\}}$ has at most one non-trivial Jordan block, $X$ can never be contained in any of them; thus, the orbit of $X$ is associated to $\fl_{\{\alpha_1,\alpha_3\}}$ and has the Bala--Carter label $2A_1$.\\
\end{example}

\subsection{Physical Origin of Bala--Carter Labels}

The relation to the little string defect classification of section \ref{sec:review} is immediate:
since polarized defects of the little string distinguish the parabolic subalgebra $\fp_{\Theta}$ of $\fg$ in the CFT limit, we simply identify the set of simple roots $\Theta$ with the Bala--Carter label of the defect. Namely, the union of all the elements of $\Theta$ forms a subquiver of $\fg$, which denotes the Bala--Carter label for the defect. The corresponding nilpotent orbit is the principal nilpotent orbit of $\fl_{\Theta}$, defined in the previous section. Equivalently, the Bala--Carter label is given by the union of the simple roots $\alpha_i$ in the Toda level 1 null state condition:
\[
\vec\beta\cdot\vec\alpha_i=0 \quad \forall \vec\alpha_i\in\Theta,
\]
for some highest weight state $|\vec\beta\rangle$ of the  $\cW(\fg)$-algebra. See figure \ref{fig:BalaCarterPolarized} below.

If the polarized theory $T^{2d}$ is described by the Bala--Carter label denoting a nilpotent orbit $\cO$, its Coulomb branch is a resolution of the Spaltenstein dual of $\cO$.\\

\begin{figure}[htpb]
	\begin{center}
	\begin{tabular}{cc}
	\large$\fg$&Bala--Carter Classification and Polarized Little String Defects\\
	\toprule
	$A_3$&
	        \tcbset{enhanced,size=fbox,nobeforeafter,tcbox raise=-2.3mm,colback=white,colframe=white}
		\begin{tikzpicture}[baseline,font=\small]
		\node at (-0.5,0) {$\Theta=\{\alpha_2,\alpha_3\}$};
		\draw[->, -stealth,  line width=0.4em, postaction={draw,-stealth,white,line width=0.2em,
                shorten <=0.10em,shorten >=0.26em}](2,0) -- (1,0);
		\node[align=justify] at (4,0) {$\omega_1:[\phantom{-}1,\phantom{-}0,\phantom{-}0]$ \\ $\omega_2:[-1,\phantom{-}0,\phantom{-}0]$};
		\draw[->, -stealth,  line width=0.4em](6,0) -- (7,0);
		\node at (9,0) {\begin{tikzpicture}
 \begin{scope}[auto, every node/.style={minimum size=0.75cm}]
\def \spac {1cm}

\node[circle, draw](k1) at (0,0) {$1$};
\node[circle, draw,red](k2) at (1*\spac,0) {$1$};
\node[circle, draw,red](k3) at (2*\spac,0) {$1$};

\node[draw, inner sep=0.1cm,minimum size=0.67cm](N1) at (0*\spac,\spac) {$1$};
\node[draw, inner sep=0.1cm,minimum size=0.67cm](N3) at (2*\spac,\spac) {$1$};

\draw[-] (k1) to (k2);
\draw[-,red] (k2) to (k3);

\draw (k1) -- (N1);
\draw (k3) -- (N3);

\end{scope}
\end{tikzpicture}};
\node[text width=12em,align=center] at (-0.5,-0.7) {Bala--Carter label: {\color{red!70!black}$A_2$}};
		\end{tikzpicture}\\
			$D_4$&
	        \tcbset{enhanced,size=fbox,nobeforeafter,tcbox raise=-2.3mm,colback=white,colframe=white}
		\begin{tikzpicture}[baseline,font=\small]
		\node at (-0.5,0) {$\Theta=\{\alpha_2,\alpha_3,\alpha_4\}$};
		\draw[->, -stealth,  line width=0.4em, postaction={draw,-stealth,white,line width=0.2em,
                shorten <=0.10em,shorten >=0.26em}](2,0) -- (1,0);
		\node[align=justify] at (4,0) {$\omega_1:[\phantom{-}1,\phantom{-}0,\phantom{-}0,\phantom{-}0]$ \\ $\omega_2:[-1,\phantom{-}0,\phantom{-}0,\phantom{-}0]$};
		\draw[->, -stealth,  line width=0.4em](6,0) -- (7,0);
		\node at (9,0) {\begin{tikzpicture}
 \begin{scope}[auto, every node/.style={minimum size=0.75cm}]
\def \spac {1cm}

\node[circle, draw](k1) at (0,0) {$2$};
\node[circle, draw,red](k2) at (1*\spac,0) {$2$};
\node[circle, draw,red](k3) at (2*\spac,0.6*\spac) {$1$};
\node[circle, draw,red](k4) at (2*\spac,-0.6*\spac) {$1$};

\node[draw, inner sep=0.1cm,minimum size=0.67cm](N1) at (0*\spac,\spac) {$2$};

\draw[-] (k1) to (k2);
\draw[-,red] (k2) to (k3);
\draw[-,red] (k2) to (k4);

\draw (k1) -- (N1);

\end{scope}
\end{tikzpicture}};
\node[text width=12em,align=center] at (-0.5,-0.7) {Bala--Carter label: {\color{red!70!black}$A_3$}};
		\end{tikzpicture}\\
					$E_6$&
	        \tcbset{enhanced,size=fbox,nobeforeafter,tcbox raise=-2.3mm,colback=white,colframe=white}
		\begin{tikzpicture}[baseline,font=\small]
		\node at (-0.5,0) {$\Theta=\{\alpha_2,\alpha_3,\alpha_4,\alpha_6\}$};
		\draw[->, -stealth,  line width=0.4em, postaction={draw,-stealth,white,line width=0.2em,
                shorten <=0.10em,shorten >=0.26em}](2.1,0) -- (1.2,0);
		\node[align=justify] at (4,0) {$\omega_1:[\phantom{-}0,0,0,0,-1,0]$ \\ $\omega_2:[-1,0,0,0,\phantom{-}1,0]$\\$\omega_3:[\phantom{-}1,0,0,0,\phantom{-}0,0]$};
		\draw[->, -stealth,  line width=0.4em](5.9,0) -- (6.7,0);
		\node at (9,0) {\begin{tikzpicture}
 \begin{scope}[auto, every node/.style={minimum size=0.75cm}]
\def \spac {1cm}

\node[circle, draw](k1) at (0,0) {$2$};
\node[circle, draw,red](k2) at (1*\spac,0) {$4$};
\node[circle, draw,red](k3) at (2*\spac,0*\spac) {$6$};
\node[circle, draw,red](k4) at (3*\spac,0*\spac) {$5$};
\node[circle, draw](k5) at (4*\spac,0*\spac) {$4$};
\node[circle, draw,red](k6) at (2*\spac,1*\spac) {$3$};

\node[draw, inner sep=0.1cm,minimum size=0.67cm](N5) at (4*\spac,\spac) {$3$};

\draw[-] (k1) to (k2);
\draw[-,red] (k2) to (k3);
\draw[-,red] (k3) to (k6);
\draw[-,red] (k3) to (k4);
\draw[-,red] (k4) to (k5);

\draw (k5) -- (N5);

\end{scope}
\end{tikzpicture}};
\node[text width=12em,align=center] at (-0.5,-0.7) {Bala--Carter label: {\color{red!70!black}$D_4$}};
		\end{tikzpicture}
	\end{tabular}
	\end{center}
	\caption{From a distinguished set of weights $\cW_{\cS}$ defining a polarized little string defect theory $T^{2d}$, one can extract a quiver gauge theory, shown on the right, and a parabolic subalgebra $\fp_{\Theta}$ in the  $m_s\rightarrow\infty$ limit, shown on the left. The set $\Theta$ defines a Bala--Carter label, also shown in red as a subquiver of $\fg$ on the right. Note $\vec\beta\cdot\vec\alpha_i=0$ for all $\vec\alpha_i\in\Theta$, which defines a null state at level 1 in $\fg$-Toda.}
	\label{fig:BalaCarterPolarized}
\end{figure}
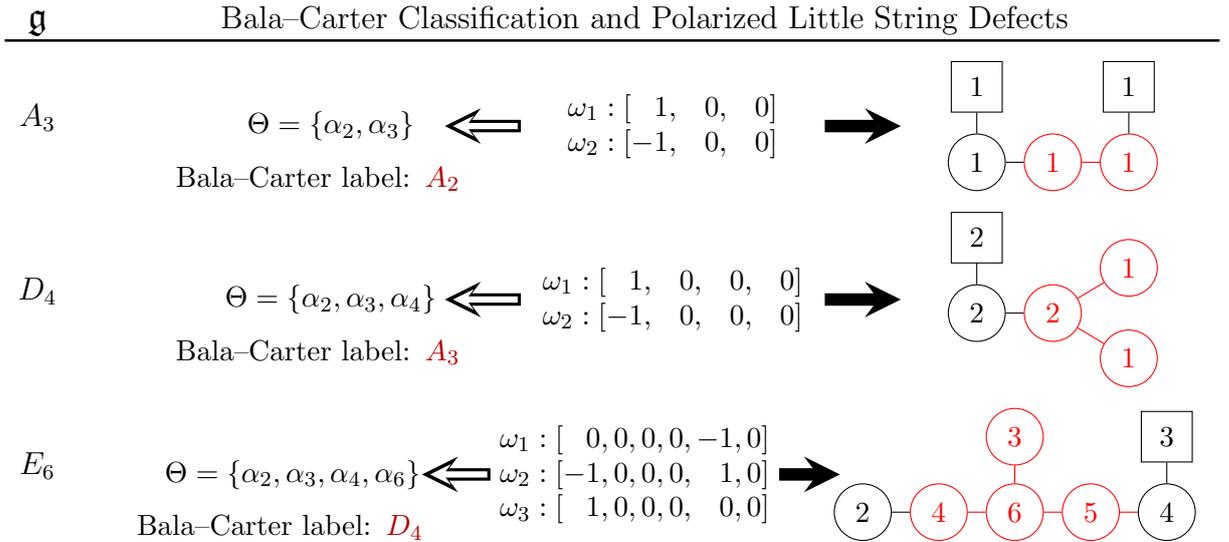


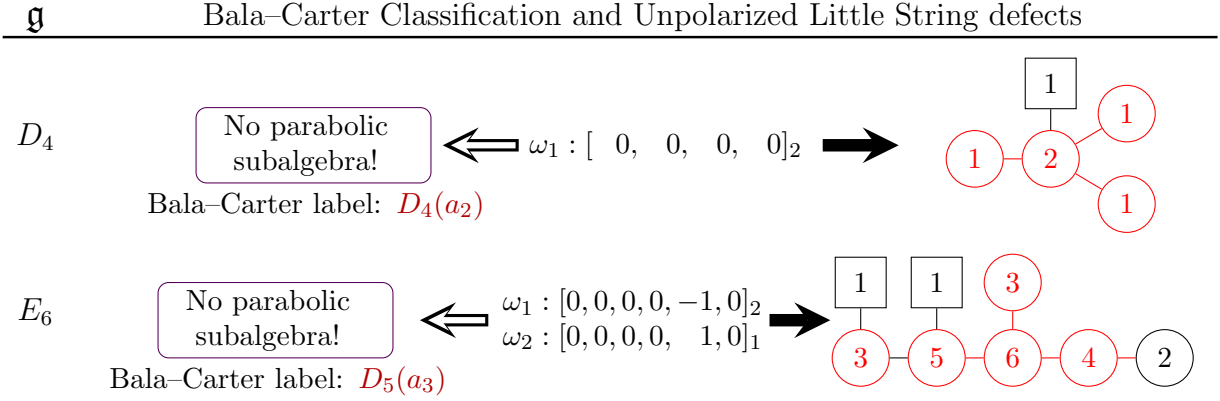
\begin{figure}[htpb]
	\begin{center}
\begin{tabular}{cc}
	\large$\fg$&Bala--Carter Classification and Unpolarized Little String defects\\
	\toprule
	$D_4$&
	        \tcbset{enhanced,size=fbox,nobeforeafter,tcbox raise=-2.3mm,colback=white,colframe=white}
		\begin{tikzpicture}[baseline,font=\small]
		\node[draw, text width=7em, align=center,rectangle, rounded corners, inner sep=0.1cm, color=violet!70!black, text=black] at (-0.7,0) {No parabolic subalgebra!};
		
		\draw[->, -stealth,  line width=0.4em, postaction={draw,-stealth,white,line width=0.2em,
                shorten <=0.10em,shorten >=0.26em}](2,0) -- (1,0);
		\node[align=justify] at (4,0) {$\omega_1:[\phantom{-}0,\phantom{-}0,\phantom{-}0,\phantom{-}0]_2$ };
		\draw[->, -stealth,  line width=0.4em](6,0) -- (7,0);
		\node at (9,0) {\begin{tikzpicture}
 \begin{scope}[auto, every node/.style={minimum size=0.75cm}]
\def \spac {1cm}

\node[circle, draw,red](k1) at (0,0) {$1$};
\node[circle, draw,red](k2) at (1*\spac,0) {$2$};
\node[circle, draw,red](k3) at (2*\spac,0.6*\spac) {$1$};
\node[circle, draw,red](k4) at (2*\spac,-0.6*\spac) {$1$};

\node[draw, inner sep=0.1cm,minimum size=0.67cm](N2) at (1*\spac,\spac) {$1$};

\draw[-,red] (k1) to (k2);
\draw[-,red] (k2) to (k3);
\draw[-,red] (k2) to (k4);

\draw (k2) -- (N2);

\end{scope}
\end{tikzpicture}};
\node[text width=12em,align=center] at (-0.5,-0.8) {Bala--Carter label: {\color{red!70!black}$D_4(a_2)$}};
		\end{tikzpicture}\\
					$E_6$&
	        \tcbset{enhanced,size=fbox,nobeforeafter,tcbox raise=-2.3mm,colback=white,colframe=white}
		\begin{tikzpicture}[baseline,font=\small]
	\node[draw, text width=7em, align=center,rectangle, rounded corners, inner sep=0.1cm, color=violet!70!black, text=black] at (-0.7,0) {No parabolic subalgebra!};
		\draw[->, -stealth,  line width=0.4em, postaction={draw,-stealth,white,line width=0.2em,
                shorten <=0.10em,shorten >=0.26em}](2.1,0) -- (1.2,0);
		\node[align=justify] at (4,0) {$\omega_1:[0,0,0,0,-1,0]_2$ \\ $\omega_2:[0,0,0,0,\phantom{-}1,0]_1$};
		\draw[->, -stealth,  line width=0.4em](5.8,0) -- (6.6,0);
		\node at (9,0) {\begin{tikzpicture}
 \begin{scope}[auto, every node/.style={minimum size=0.75cm}]
\def \spac {1cm}

\node[circle, draw,red](k1) at (0,0) {$3$};
\node[circle, draw,red](k2) at (1*\spac,0) {$5$};
\node[circle, draw,red](k3) at (2*\spac,0*\spac) {$6$};
\node[circle, draw,red](k4) at (3*\spac,0*\spac) {$4$};
\node[circle, draw](k5) at (4*\spac,0*\spac) {$2$};
\node[circle, draw,red](k6) at (2*\spac,1*\spac) {$3$};

\node[draw, inner sep=0.1cm,minimum size=0.67cm](N1) at (0*\spac,\spac) {$1$};
\node[draw, inner sep=0.1cm,minimum size=0.67cm](N2) at (1*\spac,\spac) {$1$};

\draw[-] (k1) to (k2);
\draw[-,red] (k2) to (k3);
\draw[-,red] (k3) to (k6);
\draw[-,red] (k3) to (k4);
\draw[-,red] (k4) to (k5);

\draw (k1) -- (N1);
\draw (k2) -- (N2);

\end{scope}
\end{tikzpicture}};
\node[text width=12em,align=center] at (-0.5,-0.8) {Bala--Carter label: {\color{red!70!black}$D_5(a_3)$}};
		\end{tikzpicture}
	\end{tabular}
	\end{center}
	\caption{From a distinguished set of weights $\cW_{\cS}$ defining an unpolarized little string defect theory $T^{2d}$, one can extract a quiver gauge theory, shown on the right, but no parabolic subalgebra $\fp_{\Theta}$ in the  $m_s\rightarrow\infty$ limit, as shown on the left. We added a subscript denoting which representation the weights $\omega_i$ belong in to fully specify the Bala--Carter label. The additional simple root data of the Bala--Carter label is written as $a_i$, where $i$ is a number of simple roots.}
	\label{fig:BalaCarterUnpolarized}
\end{figure}


Concerning unpolarized defects of the little string, recall that they are characterized as follows:  either  $\cW_{\cS}$ is the set containing the zero weight $\omega=[0,0,\ldots,0]$ only (possibly multiple times), or $\cW_{\cS}$ contains a nonzero weight $\omega$ in the representation generated by (minus) a fundamental weight $-w_a$ without being in the Weyl orbit of $-w_a$.

Either way, additional data is needed to characterize such defects: in the end, it is sufficient to specify the representation $\omega$ belongs in.  This prescription is in one-to-one correspondence with specifying a set of additional simple roots next to the Bala--Carter label of a non-principal nilpotent orbit, as we explained in  section \ref{ssec:balalabel}. To our knowledge, this extra simple root label unfortunately does not have a nice geometric interpretation for the defect. See figure \ref{fig:BalaCarterUnpolarized} for examples.

At any rate, note that an unpolarized defect will still satisfy the relation
\[
\vec\beta\cdot\vec\alpha_i=0 \quad \forall \vec\alpha_i\in\Theta,
\]
for some subset of positive simple roots $\Theta$, with $\vec\beta=\sum_{i=1}^{|\cW_{\cS}|}\beta_i\; \omega_i$. This is the same level 1 null state condition of $\fg$-type Toda satisfied by polarized defects. There is however one crucial difference: the above constraint is no longer sufficient to characterize the defect, and one should specify the representation each $\omega_i$ belongs in.\\

\begin{example}
In the case of the single zero weight $\omega=[0,0,\ldots,0]$, we get $\vec\beta=0$ and the null state condition is of course trivially satisfied; however, one should also specify which representation the weight $\omega$ is taken in, since for a given algebra $\fg$, $\omega$ belongs in general to many representations. This corresponds to specifying additional simple roots next to the Bala--Carter label $\fg$. Note the Bala--Carter label $\fg$ without any extra simple roots specified denotes the trivial nilpotent orbit, that is to say the absence of a defect.\\
\end{example}

In this way, one derives the full Bala--Carter classification of nilpotent orbits simply from a distinguished set $\cW_{\cS}$ of weights defining a little string defect. It would be interesting to extend the analysis to the non-simply laced semi-simple Lie algebras, for which a Bala--Carter classification is also available. We leave the study of these defects to future work.

\section{Weighted Dynkin Diagrams}
\label{sec:dynkin}

There is yet another way to classify nilpotent orbits of $\fg$, known as the so-called \emph{weighted Dynkin diagrams.} We now show how to derive them.

\subsection{Mathematical construction}
Weighted Dynkin diagrams are vectors of integers $r_i\in\{0,1,2\}$, where $i=1,\ldots , \text{rk}\fg$; thus, we get one number for each node in the Dynkin diagram of $\fg$. We can associate such a vector to each nilpotent orbit of $\fg$, and each nilpotent orbit has a unique weighted Dynkin diagram. Note, however, that not all such labellings of the Dynkin diagram also have a nilpotent orbit corresponding to it.\\

To  construct such a weighted Dynkin diagram, we use the following theorem by Jacobson and Morozov \cite{Morozov:1942,*Jacobson:1951}. 

Remember that $\mathfrak{sl}_2$ is the algebra generated by $X, Y$ and $H$ with the relations
\begin{equation}
[H,X]=2X,\qquad [H,Y]=-2Y, \qquad [X,Y]=H.
\end{equation}

Every nilpotent orbit in $\fg$ arises as the orbit of the image of $X$ in an embedding $\rho:\mathfrak{sl}_2\to\fg$.

In other words, for any embedding $\rho:\mathfrak{sl}_2\to\fg$, the element $\rho(X)$ always is a nilpotent element of $\fg$. The Jacobson--Morozov theorem tells us that any nilpotent orbit uniquely arises (up to conjugation) as the orbit of such an element.

This means in particular that any nilpotent orbit also determines an element $\rho(H)$, which is semi-simple (we assume it to be diagonal). For simplicity, we'll just write $\rho(H)$ as $H$. The (diagonal) entries of $H$ are always integers, and allow us to read off the weighted Dynkin diagram; the entry of the $i$-th node is defined to be $r_i=\alpha_i(H)$, where $\alpha_i$ is the $i$-th simple root of $\fg$. It turns out that these numbers are always $0$, $1$ or $2$.\\

\begin{example}
We illustrate the above construction for the nilpotent orbit of

$$
X=\begin{pmatrix}
0&1&0&0\\
0&0&0&0\\
0&0&0&1\\
0&0&0&0\\
\end{pmatrix}
$$

in $\mathfrak{sl}_4$. One first constructs $H$; we won't do this explicitly here (see \cite{Collingwood:1993} for details), but the result is

$$
H=\begin{pmatrix}
1&0&0&0\\
0&-1&0&0\\
0&0&1&0\\
0&0&0&-1\\
\end{pmatrix}.
$$
The next step is to reorder the elements in the diagonal of $H$ in a monotonically decreasing order. The quadruple we get is $(h_1,h_2,h_3,h_4)=(1,1,-1,-1)$.

The nodes of the Dynkin diagram are labelled by the consecutive differences of these numbers, so $r_i=h_i-h_{i+1}$. This gives us $(r_1,r_2,r_3)=(0,2,0)$. So the weighted Dynkin diagram in this example looks as follows:

\begin{center}
\begin{tikzpicture}[baseline,font=\footnotesize]
 \begin{scope}[auto, every node/.style={minimum size=0.35cm}]
\def \spac {0.85cm}

\node[rectangle](k1) at (0,0) {$0$};
\node[rectangle](k2) at (1*\spac,0) {$2$};
\node[rectangle](k3) at (2*\spac,0*\spac) {$0$};

\draw[-] (k1) to (k2);
\draw[-] (k2) to (k3);

\end{scope}
\end{tikzpicture}
\end{center}
\end{example}

One can generalize the above construction to all semi-simple Lie algebras, with minor modifications.

\subsection{From Weighted Dynkin Diagrams to Little String Defects}

We make the following  observations:

All weighted Dynkin diagrams can be interpreted as physical quiver theories: the label on each node of the weighted Dynkin diagram should be understood as the rank of a flavor symmetry group in a quiver. The  quivers one reads  in this way are always superconformal (in a 4d sense), and the flavor symmetry on each node is either nothing, a $U(1)$ group, or a $U(2)$ group. For instance, the full puncture, or maximal nilpotent orbit, denoted by the weighted Dynkin diagram $(2,2,\ldots,2,2)$, can be understood as a quiver gauge theory with a  $U(2)$ flavor attached to each node, for all semi-simple Lie algebras (see also \cite{Hanany:2016gbz}). Pushing this idea further, we find, surprisingly, that these quivers are little string defect theories $T^{2d}$, at finite $m_s$.\\

\begin{figure}[h!]
\begin{tabular}{c}
\begin{tikzpicture}[baseline,font=\small]
	\node at (-2.5,0) {\begin{tikzpicture}[baseline,font=\footnotesize]
 \begin{scope}[auto, every node/.style={minimum size=0.65cm}]
\def \spac {0.9cm}

\node[circle, draw](k1) at (0,0) {$1$};
\node[circle, draw](k2) at (1*\spac,0) {$1$};
\node[circle, draw](k3) at (2*\spac,0*\spac) {$1$};
\node[circle, draw](k4) at (3*\spac,0*\spac) {$1$};

\node[draw, inner sep=0.1cm,minimum size=0.55cm](N1) at (0*\spac,1*\spac) {$1$};
\node[draw, inner sep=0.1cm,minimum size=0.55cm](N4) at (3*\spac,1*\spac) {$1$};

\draw[-] (k1) to (k2);
\draw[-] (k2) to (k3);
\draw[-] (k3) to (k4);

\draw (k1) -- (N1);
\draw (k4) -- (N4);

\end{scope}
\end{tikzpicture}};
		\node[align=justify, font=\small] at (5.3,-0.35) {$(1,0,0,1)$};
		\draw[->, -stealth,  line width=0.3em](-0.35,-0.35) -- (0.35,-0.35);
		\node at (2.5,0) {\begin{tikzpicture}[baseline,font=\footnotesize]
 \begin{scope}[auto, every node/.style={minimum size=0.65cm}]
\def \spac {0.9cm}

\node[circle, draw](k1) at (0,0) {$1$};
\node[circle, draw](k2) at (1*\spac,0) {$1$};
\node[circle, draw](k3) at (2*\spac,0*\spac) {$1$};
\node[circle, draw](k4) at (3*\spac,0*\spac) {$1$};

\node[draw, inner sep=0.1cm,minimum size=0.55cm](N1) at (0*\spac,1*\spac) {$1$};
\node[draw, inner sep=0.1cm,minimum size=0.55cm](N4) at (3*\spac,1*\spac) {$1$};

\draw[-] (k1) to (k2);
\draw[-] (k2) to (k3);
\draw[-] (k3) to (k4);

\draw (k1) -- (N1);
\draw (k4) -- (N4);

\end{scope}
\end{tikzpicture}};
		\end{tikzpicture}\\
		

\begin{tikzpicture}[baseline,font=\small]
	\node at (-2.5,0) {\begin{tikzpicture}[baseline,font=\footnotesize]
 \begin{scope}[auto, every node/.style={minimum size=0.65cm}]
\def \spac {0.9cm}

\node[circle, draw](k1) at (0,0) {$1$};
\node[circle, draw](k2) at (1*\spac,0) {$2$};
\node[circle, draw](k3) at (2*\spac,0*\spac) {$2$};
\node[circle, draw](k4) at (3*\spac,0*\spac) {$1$};

\node[draw, inner sep=0.1cm,minimum size=0.55cm](N2) at (1*\spac,1*\spac) {$1$};
\node[draw, inner sep=0.1cm,minimum size=0.55cm](N3) at (2*\spac,1*\spac) {$1$};

\draw[-] (k1) to (k2);
\draw[-] (k2) to (k3);
\draw[-] (k3) to (k4);

\draw (k2) -- (N2);
\draw (k3) -- (N3);

\end{scope}
\end{tikzpicture}};
		\node[align=justify, font=\small] at (5.3,-0.35) {$(0,1,1,0)$};
		\draw[->, -stealth,  line width=0.3em](-0.35,-0.35) -- (0.35,-0.35);
		\node at (2.5,0) {\begin{tikzpicture}[baseline,font=\footnotesize]
 \begin{scope}[auto, every node/.style={minimum size=0.65cm}]
\def \spac {0.9cm}

\node[circle, draw](k1) at (0,0) {$1$};
\node[circle, draw](k2) at (1*\spac,0) {$2$};
\node[circle, draw](k3) at (2*\spac,0*\spac) {$2$};
\node[circle, draw](k4) at (3*\spac,0*\spac) {$1$};

\node[draw, inner sep=0.1cm,minimum size=0.55cm](N2) at (1*\spac,1*\spac) {$1$};
\node[draw, inner sep=0.1cm,minimum size=0.55cm](N3) at (2*\spac,1*\spac) {$1$};

\draw[-] (k1) to (k2);
\draw[-] (k2) to (k3);
\draw[-] (k3) to (k4);

\draw (k2) -- (N2);
\draw (k3) -- (N3);

\end{scope}
\end{tikzpicture}};
		\end{tikzpicture}\\
				

\begin{tikzpicture}[baseline,font=\small]
	\node at (-2.5,0) {\begin{tikzpicture}[baseline,font=\footnotesize]
 \begin{scope}[auto, every node/.style={minimum size=0.65cm}]
\def \spac {0.9cm}

\node[circle, draw](k1) at (0,0) {$2$};
\node[circle, draw](k2) at (1*\spac,0) {$2$};
\node[circle, draw](k3) at (2*\spac,0*\spac) {$2$};
\node[circle, draw](k4) at (3*\spac,0*\spac) {$1$};

\node[draw, inner sep=0.1cm,minimum size=0.55cm](N1) at (0*\spac,1*\spac) {$2$};
\node[draw, inner sep=0.1cm,minimum size=0.55cm](N3) at (2*\spac,1*\spac) {$1$};

\draw[-] (k1) to (k2);
\draw[-] (k2) to (k3);
\draw[-] (k3) to (k4);

\draw (k1) -- (N1);
\draw (k3) -- (N3);

\end{scope}
\end{tikzpicture}};
		\node[align=justify, font=\small] at (5.3,-0.35) {$(2,0,0,2)$};
		\node[align=justify, font=\scriptsize] at (0,0.05) {H. W.};
		\draw[->, -stealth,  line width=0.3em](-0.35,-0.35) -- (0.35,-0.35);
		\node at (2.5,0) {\begin{tikzpicture}[baseline,font=\footnotesize]
 \begin{scope}[auto, every node/.style={minimum size=0.65cm}]
\def \spac {0.9cm}

\node[circle, draw](k1) at (0,0) {$2$};
\node[circle, draw](k2) at (1*\spac,0) {$2$};
\node[circle, draw](k3) at (2*\spac,0*\spac) {$2$};
\node[circle, draw](k4) at (3*\spac,0*\spac) {$2$};

\node[draw, inner sep=0.1cm,minimum size=0.55cm](N1) at (0*\spac,1*\spac) {$2$};
\node[draw, inner sep=0.1cm,minimum size=0.55cm](N4) at (3*\spac,1*\spac) {$2$};

\draw[-] (k1) to (k2);
\draw[-] (k2) to (k3);
\draw[-] (k3) to (k4);

\draw (k1) -- (N1);
\draw (k4) -- (N4);

\end{scope}
\end{tikzpicture}};
		\end{tikzpicture}\\


\begin{tikzpicture}[baseline,font=\small]
	\node at (-2.5,0) {\begin{tikzpicture}[baseline,font=\footnotesize]
 \begin{scope}[auto, every node/.style={minimum size=0.65cm}]
\def \spac {0.9cm}

\node[circle, draw](k1) at (0,0) {$2$};
\node[circle, draw](k2) at (1*\spac,0) {$3$};
\node[circle, draw](k3) at (2*\spac,0*\spac) {$2$};
\node[circle, draw](k4) at (3*\spac,0*\spac) {$1$};

\node[draw, inner sep=0.1cm,minimum size=0.55cm](N1) at (0*\spac,1*\spac) {$1$};
\node[draw, inner sep=0.1cm,minimum size=0.55cm](N2) at (1*\spac,1*\spac) {$2$};

\draw[-] (k1) to (k2);
\draw[-] (k2) to (k3);
\draw[-] (k3) to (k4);

\draw (k1) -- (N1);
\draw (k2) -- (N2);

\end{scope}
\end{tikzpicture}};
		\node[align=justify, font=\scriptsize] at (0,0.05) {H. W.};
		\node[align=justify, font=\small] at (5.3,-0.35) {$(1,1,1,1)$};
		\draw[->, -stealth,  line width=0.3em](-0.35,-0.35) -- (0.35,-0.35);
		\node at (2.5,0) {\begin{tikzpicture}[baseline,font=\footnotesize]
 \begin{scope}[auto, every node/.style={minimum size=0.65cm}]
\def \spac {0.9cm}

\node[circle, draw](k1) at (0,0) {$2$};
\node[circle, draw](k2) at (1*\spac,0) {$3$};
\node[circle, draw](k3) at (2*\spac,0*\spac) {$3$};
\node[circle, draw](k4) at (3*\spac,0*\spac) {$2$};

\node[draw, inner sep=0.1cm,minimum size=0.55cm](N1) at (0*\spac,1*\spac) {$1$};
\node[draw, inner sep=0.1cm,minimum size=0.55cm](N2) at (1*\spac,1*\spac) {$1$};
\node[draw, inner sep=0.1cm,minimum size=0.55cm](N3) at (2*\spac,1*\spac) {$1$};
\node[draw, inner sep=0.1cm,minimum size=0.55cm](N4) at (3*\spac,1*\spac) {$1$};

\draw[-] (k1) to (k2);
\draw[-] (k2) to (k3);
\draw[-] (k3) to (k4);

\draw (k1) -- (N1);
\draw (k2) -- (N2);
\draw (k3) -- (N3);
\draw (k4) -- (N4);

\end{scope}
\end{tikzpicture}};
		\end{tikzpicture}\\
		

\begin{tikzpicture}[baseline,font=\small]
	\node at (-2.5,0) {\begin{tikzpicture}[baseline,font=\footnotesize]
 \begin{scope}[auto, every node/.style={minimum size=0.65cm}]
\def \spac {0.9cm}

\node[circle, draw](k1) at (0,0) {$3$};
\node[circle, draw](k2) at (1*\spac,0) {$3$};
\node[circle, draw](k3) at (2*\spac,0*\spac) {$2$};
\node[circle, draw](k4) at (3*\spac,0*\spac) {$1$};

\node[draw, inner sep=0.1cm,minimum size=0.55cm](N1) at (0*\spac,1*\spac) {$3$};
\node[draw, inner sep=0.1cm,minimum size=0.55cm](N2) at (1*\spac,1*\spac) {$1$};

\draw[-] (k1) to (k2);
\draw[-] (k2) to (k3);
\draw[-] (k3) to (k4);

\draw (k1) -- (N1);
\draw (k2) -- (N2);

\end{scope}
\end{tikzpicture}};
		\node[align=justify, font=\scriptsize] at (0,0.05) {H. W.};
		\node[align=justify, font=\small] at (5.3,-0.35) {$(2,1,1,2)$};
		\draw[->, -stealth,  line width=0.3em](-0.35,-0.35) -- (0.35,-0.35);
		\node at (2.5,0) {\begin{tikzpicture}[baseline,font=\footnotesize]
 \begin{scope}[auto, every node/.style={minimum size=0.65cm}]
\def \spac {0.9cm}

\node[circle, draw](k1) at (0,0) {$3$};
\node[circle, draw](k2) at (1*\spac,0) {$4$};
\node[circle, draw](k3) at (2*\spac,0*\spac) {$4$};
\node[circle, draw](k4) at (3*\spac,0*\spac) {$3$};

\node[draw, inner sep=0.1cm,minimum size=0.55cm](N1) at (0*\spac,1*\spac) {$2$};
\node[draw, inner sep=0.1cm,minimum size=0.55cm](N2) at (1*\spac,1*\spac) {$1$};
\node[draw, inner sep=0.1cm,minimum size=0.55cm](N3) at (2*\spac,1*\spac) {$1$};
\node[draw, inner sep=0.1cm,minimum size=0.55cm](N4) at (3*\spac,1*\spac) {$2$};

\draw[-] (k1) to (k2);
\draw[-] (k2) to (k3);
\draw[-] (k3) to (k4);

\draw (k1) -- (N1);
\draw (k2) -- (N2);
\draw (k3) -- (N3);
\draw (k4) -- (N4);

\end{scope}
\end{tikzpicture}};
		\end{tikzpicture}\\


\begin{tikzpicture}[baseline,font=\small]
	\node at (-2.5,0) {\begin{tikzpicture}[baseline,font=\footnotesize]
 \begin{scope}[auto, every node/.style={minimum size=0.65cm}]
\def \spac {0.9cm}

\node[circle, draw](k1) at (0,0) {$4$};
\node[circle, draw](k2) at (1*\spac,0) {$3$};
\node[circle, draw](k3) at (2*\spac,0*\spac) {$2$};
\node[circle, draw](k4) at (3*\spac,0*\spac) {$1$};

\node[draw, inner sep=0.1cm,minimum size=0.55cm](N1) at (0*\spac,1*\spac) {$5$};

\draw[-] (k1) to (k2);
\draw[-] (k2) to (k3);
\draw[-] (k3) to (k4);

\draw (k1) -- (N1);

\end{scope}
\end{tikzpicture}};
		\node[align=justify, font=\scriptsize] at (0,0.05) {H. W.};
		\node[align=justify, font=\small] at (5.3,-0.35) {$(2,2,2,2)$};
		\draw[->, -stealth,  line width=0.3em](-0.35,-0.35) -- (0.35,-0.35);
		\node at (2.5,0) {\begin{tikzpicture}[baseline,font=\footnotesize]
 \begin{scope}[auto, every node/.style={minimum size=0.65cm}]
\def \spac {0.9cm}

\node[circle, draw](k1) at (0,0) {$4$};
\node[circle, draw](k2) at (1*\spac,0) {$6$};
\node[circle, draw](k3) at (2*\spac,0*\spac) {$6$};
\node[circle, draw](k4) at (3*\spac,0*\spac) {$4$};

\node[draw, inner sep=0.1cm,minimum size=0.55cm](N1) at (0*\spac,1*\spac) {$2$};
\node[draw, inner sep=0.1cm,minimum size=0.55cm](N2) at (1*\spac,1*\spac) {$2$};
\node[draw, inner sep=0.1cm,minimum size=0.55cm](N3) at (2*\spac,1*\spac) {$2$};
\node[draw, inner sep=0.1cm,minimum size=0.55cm](N4) at (3*\spac,1*\spac) {$2$};

\draw[-] (k1) to (k2);
\draw[-] (k2) to (k3);
\draw[-] (k3) to (k4);

\draw (k1) -- (N1);
\draw (k2) -- (N2);
\draw (k3) -- (N3);
\draw (k4) -- (N4);

\end{scope}
\end{tikzpicture}};
		\end{tikzpicture}
\end{tabular}
	\caption{Either directly, or after flowing on the Higgs branch by Hanany-Witten transition to symmetrize the theories $T^{2d}$, the little string quivers (left) are precisely the weighted Dynkin diagrams of $\fg$ (right); the integers $0, 1, 2$ then get an interpretation as flavor symmetry ranks. Shown above is the case $\fg = A_4$.}
	\label{fig:jacobsonA4}
\end{figure}
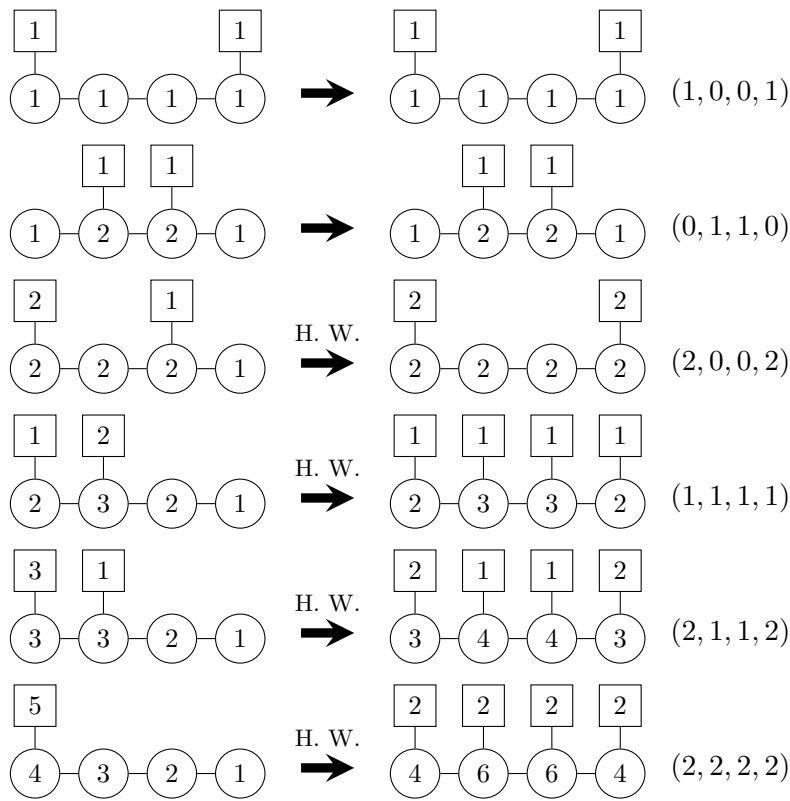

In the case of $\fg=A_n$, this correspondence between weighted Dynkin diagrams and defect theories $T^{2d}$ can be made explicit.
Indeed, all $A_n$ weighted Dynkin diagrams are invariant under the $\mathbb{Z}_2$ outer automorphism action of the algebra; in other words, the quivers are all symmetric. For low dimensional defects, these quivers are precisely the little string quivers $T^{2d}$ studied in this note. For instance, consider the simple puncture of $A_n$, generated by the set of weights $\cW_{\cS}=\{[1,0,\ldots,0] \; ,\; [-1,0,\ldots,0] \}$, with Bala-Carter label $A_{n-1}$; the weighted Dynkin diagram with this Bala-Carter label can be shown to be $(1,0,\ldots,0,1)$, in standard notation. This is precisely the little string quiver $T^{2d}$ for the simple puncture! It has a $U(1)$ flavor symmetry on the first node, and a $U(1)$ flavor symmetry on the last node, as it should. See figure \ref{fig:jacobsonA4}.

Many of the little string quivers $T^{2d}$ of $A_n$, however, are not weighted Dynkin diagrams. They are the quivers not invariant under $\mathbb{Z}_2$ reflection. We claim that such theories $T^{2d}$ can however uniquely be turned into the correct weighted Dynkin diagrams, by moving on the Higgs branch of the theories.

\begin{figure}[thpb]
	\begin{center}
		\begin{tikzpicture}[font=\small]
	\node at (-3.5,0) {\begin{tikzpicture}[baseline]
\draw [-] (0,1) -- (0,-1);
\draw [-] (0.5,1) -- (0.5,-1);
\draw [-] (1.0,1) -- (1.0,-1);
\draw [-] (1.5,1) -- (1.5,-1);

\draw[-,green!90!black,thick] (0.0,-0.4)--(0.5,-0.4);
\draw[-,green!90!black,thick] (1.0,-0.4)--(1.5,-0.4);
\draw[-,green!90!black,thick] (0.5,-0.4)--(1.0,-0.4);
\draw[-,green!90!black,thick] (0.5,-0.3)--(1.0,-0.3);

\draw (0.75,0.6) node[cross=4pt,red,ultra thick]{};
\draw (0.75,-0.35) node[cross=4pt,green!90!black,ultra thick]{};

\node[align=left, font=\tiny] at (0.75,0.80) {$\omega_2$};
\node[align=left, font=\tiny] at (0.75,-0.10) {$\omega_1$};
\end{tikzpicture}
};
	\node at (3.5,0) {
	\begin{tikzpicture}[baseline]
\draw [-] (0,1) -- (0,-1);
\draw [-] (0.5,1) -- (0.5,-1);
\draw [-] (1.0,1) -- (1.0,-1);
\draw [-] (1.5,1) -- (1.5,-1);

\draw[-,green!90!black,thick] (0.0,-0.4)--(0.5,-0.4);
\draw[-,green!90!black,thick] (1.0,-0.4)--(1.5,-0.4);
\draw[-,green!90!black,thick] (0.5,-0.4)--(1.0,-0.4);
\draw[-,green!90!black,thick] (0.5,-0.3)--(1.0,-0.3);

\draw[-,red,thick] (-0.5,0.68)--(0.5,0.68);
\draw[-,red,thick] (-0.5,0.52)--(0.0,0.52);

\draw (-0.5,0.68) node[cross=4pt,red,ultra thick]{};
\draw (-0.5,0.52) node[cross=4pt,red,ultra thick]{};
\draw (0.75,-0.35) node[cross=4pt,green!90!black,ultra thick]{};

\node[align=left, font=\tiny] at (-0.80,0.8) {$\omega_2^{\prime\prime}$};
\node[align=left, font=\tiny] at (-0.80,0.40) {$\omega_2^{\prime}$};
\node[align=left, font=\tiny] at (0.75,-0.10) {$\omega_1$};
\end{tikzpicture}
};
\end{tikzpicture}

\begin{tikzpicture}[baseline,font=\small]
	\node at (-3.75,0) {\begin{tikzpicture}[baseline,font=\footnotesize]
 \begin{scope}[auto, every node/.style={minimum size=0.65cm}]
\def \spac {0.9cm}

\node[circle, draw](k1) at (0,0) {$1$};
\node[circle, draw](k2) at (1*\spac,0) {$2$};
\node[circle, draw](k3) at (2*\spac,0*\spac) {$1$};

\node[draw, inner sep=0.1cm,minimum size=0.55cm](N2) at (1*\spac,1*\spac) {$2$};

\draw[-] (k1) to (k2);
\draw[-] (k2) to (k3);
`
\draw (k2) -- (N2);

\end{scope}
\end{tikzpicture}};
		\node[align=justify, font=\scriptsize] at (0.0,0.10) {H. W.};
		\node[align=justify, font=\footnotesize] at (-3.5,-1.4) {$\omega_1: [\phantom{-}0,\phantom{-}1,\phantom{-}0]={\color{green!50!black}-w_2+\alpha_1+2\alpha_2+\alpha_3}$\\$\omega_2: [\phantom{-}0,-1,\phantom{-}0]={\color{red!70!black}-w_2}$ };
		\node[align=justify, font=\footnotesize] at (3.7,-1.6) {$\omega_1: [\phantom{-}0,\phantom{-}1,\phantom{-}0]={\color{green!50!black}-w_2+\alpha_1+2\alpha_2+\alpha_3}$\\$\omega_2^{\prime}: [-1,\phantom{-}0,\phantom{-}0]={\color{red!70!black}-w_1}$\\ $\omega_2^{\prime\prime}: [\phantom{-}1,-1,\phantom{-}0]={\color{red!70!black}-w_1+\alpha_1}$};
\draw [decoration={brace,amplitude=0.4em},decorate,thick,gray] (0.6,-2.15) -- (0.6,-1.45);
		\draw[->, -stealth,  line width=0.4em](-0.50,-0.35) -- (0.50,-0.35);
		\node at (0.20,-1.8) {$+$};
		\node at (4.0,0) {\begin{tikzpicture}[baseline,font=\footnotesize]
 \begin{scope}[auto, every node/.style={minimum size=0.65cm}]
\def \spac {0.9cm}

\node[circle, draw](k1) at (0,0) {$2$};
\node[circle, draw](k2) at (1*\spac,0) {$2$};
\node[circle, draw](k3) at (2*\spac,0*\spac) {$1$};

\node[draw, inner sep=0.1cm,minimum size=0.55cm](N1) at (0*\spac,1*\spac) {$2$};
\node[draw, inner sep=0.1cm,minimum size=0.55cm](N2) at (1*\spac,1*\spac) {$1$};

\draw[-] (k1) to (k2);
\draw[-] (k2) to (k3);

\draw (k1) -- (N1);
\draw (k2) -- (N2);

\end{scope}
\end{tikzpicture}};
		\end{tikzpicture}
	\end{center}
	\caption{Writing a weight in a fundamental representation of $\fg$  as a sum of several weights in (possibly different) fundamental representations corresponds to flowing on the Higgs branch of $T^{2d}$. In the context of brane engineering, when $\fg=A_n$, this is the familiar Hanany--Witten transition \cite{Hanany:1996ie}. In this example, we rewrite $[0,-1,0]$ as the sum $[-1,0,0]+[1,-1,0]$. As a result, the extra Coulomb parameter $\alpha_1$ on the right is frozen to the value of the mass parameters denoted by $\omega_2'$ (and $\omega_2''$).}
	\label{fig:ExampleHananyWitten}
\end{figure}
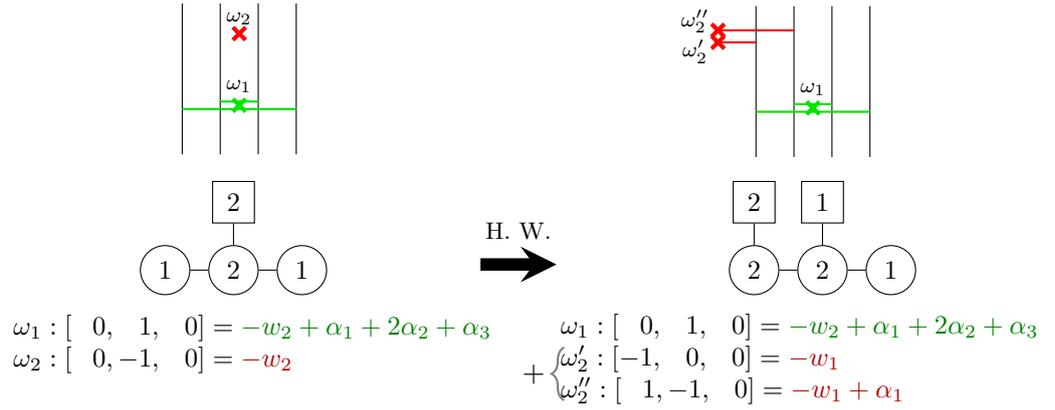

Such a flow on the Higgs branch  translates to a weight addition procedure in the algebra: this uses the fact that a weight in a fundamental representation can always be written as the sum of new weights in possibly different fundamental representations. Each of them should be in the orbit of some fundamental weight (possibly different orbits), while obeying the rule that no subset adds up to zero. In the context of brane engineering, this weight addition procedure agrees with what is referred to as Hanany--Witten transitions \cite{Hanany:1996ie}. See \cite{Haouzi:2016ohr} for details, and figure \ref{fig:ExampleHananyWitten} below for an example.
\FloatBarrier

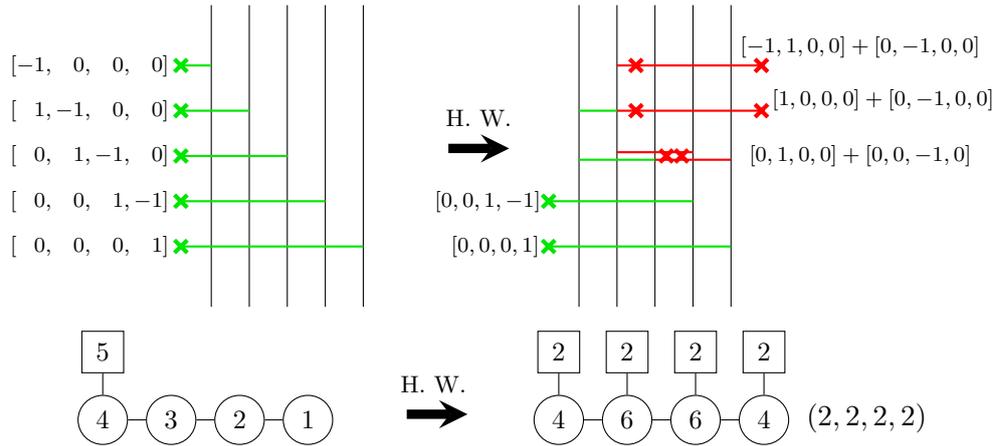
\begin{figure}[btph]
	\begin{center}
	\begin{tikzpicture}[font=\small]
	\node at (-3.5,0) {\begin{tikzpicture}[baseline]
\draw [-] (0,2) -- (0,-2);
\draw [-] (0.5,2) -- (0.5,-2);
\draw [-] (1.0,2) -- (1.0,-2);
\draw [-] (1.5,2) -- (1.5,-2);
\draw [-] (2.0,2) -- (2.0,-2);

\draw[-,green!90!black,thick] (-0.4,1.2)--(0,1.2);
\draw[-,green!90!black,thick] (-0.4,0.6)--(0.5,0.6);
\draw[-,green!90!black,thick] (-0.4,0.0)--(1.0,0.0);
\draw[-,green!90!black,thick] (-0.4,-0.6)--(1.5,-0.6);
\draw[-,green!90!black,thick] (-0.4,-1.2)--(2.0,-1.2);

\draw (-0.4,1.2) node[cross=4pt,green!90!black,ultra thick]{};
\draw (-0.4,0.6) node[cross=4pt,green!90!black,ultra thick]{};
\draw (-0.4,0.0) node[cross=4pt,green!90!black,ultra thick]{};
\draw (-0.4,-0.6) node[cross=4pt,green!90!black,ultra thick]{};
\draw (-0.4,-1.2) node[cross=4pt,green!90!black,ultra thick]{};

\node[align=left, font=\tiny] at (-1.6,1.2) {$[-1,\phantom{-}0,\phantom{-}0,\phantom{-}0]$};
\node[align=left, font=\tiny] at (-1.6,0.6) {$[\phantom{-}1,-1,\phantom{-}0,\phantom{-}0]$};
\node[align=left, font=\tiny] at (-1.6,0.0) {$[\phantom{-}0,\phantom{-}1,-1,\phantom{-}0]$};
\node[align=left, font=\tiny] at (-1.6,-0.6) {$[\phantom{-}0,\phantom{-}0,\phantom{-}1,-1]$};
\node[align=left, font=\tiny] at (-1.6,-1.2) {$[\phantom{-}0,\phantom{-}0,\phantom{-}0,\phantom{-}1]$};
\end{tikzpicture}
};
\node[align=justify, font=\scriptsize] at (0.4,0.5) {H. W.};
\draw[->, -stealth,  line width=0.3em](0.0,0.1) -- (0.8,0.1);

	\node at (3.5,0) {\begin{tikzpicture}[baseline]
\draw [-] (0,2) -- (0,-2);
\draw [-] (0.5,2) -- (0.5,-2);
\draw [-] (1.0,2) -- (1.0,-2);
\draw [-] (1.5,2) -- (1.5,-2);
\draw [-] (2.0,2) -- (2.0,-2);

\draw[-,green!90!black,thick] (0,0.6)--(0.5,0.6);
\draw[-,green!90!black,thick] (0,-0.05)--(1.0,-0.05);
\draw[-,green!90!black,thick] (-0.4,-0.6)--(1.5,-0.6);
\draw[-,green!90!black,thick] (-0.4,-1.2)--(2.0,-1.2);

\draw[-,red,thick] (0.5,1.2)--(2.4,1.2);
\draw[-,red,thick] (0.5,0.6)--(2.4,0.6);
\draw[-,red,thick] (0.5,0.05)--(1.5,0.05);
\draw[-,red,thick] (1.0,-0.05)--(2.0,-0.05);

\draw (0.75,1.2) node[cross=4pt,red,ultra thick]{};
\draw (2.4,1.2) node[cross=4pt,red,ultra thick]{};
\draw (0.75,0.6) node[cross=4pt,red,ultra thick]{};
\draw (2.4,0.6) node[cross=4pt,red,ultra thick]{};
\draw (1.15,0.0) node[cross=4pt,red,ultra thick]{};
\draw (1.35,0.0) node[cross=4pt,red,ultra thick]{};
\draw (-0.4,-0.6) node[cross=4pt,green!90!black,ultra thick]{};
\draw (-0.4,-1.2) node[cross=4pt,green!90!black,ultra thick]{};

\node[align=left, font=\tiny] at (-1.2,-0.6) {$[0,0,1,-1]$};
\node[align=left, font=\tiny] at (-1.1,-1.2) {$[0,0,0,1]$};

\node[align=left, font=\tiny] at (3.7,1.45) {$[-1,1,0,0]+[0,-1,0,0]$};
\node[align=left, font=\tiny] at (4.0,0.75) {$[1,0,0,0]+[0,-1,0,0]$};
\node[align=left, font=\tiny] at (3.7,0) {$[0,1,0,0]+[0,0,-1,0]$};
\end{tikzpicture}
};
\end{tikzpicture}
\begin{tikzpicture}[baseline,font=\small]
	\node at (-2.5,0) {\begin{tikzpicture}[baseline,font=\footnotesize]
 \begin{scope}[auto, every node/.style={minimum size=0.65cm}]
\def \spac {0.9cm}

\node[circle, draw](k1) at (0,0) {$4$};
\node[circle, draw](k2) at (1*\spac,0) {$3$};
\node[circle, draw](k3) at (2*\spac,0*\spac) {$2$};
\node[circle, draw](k4) at (3*\spac,0*\spac) {$1$};

\node[draw, inner sep=0.1cm,minimum size=0.55cm](N1) at (0*\spac,1*\spac) {$5$};

\draw[-] (k1) to (k2);
\draw[-] (k2) to (k3);
\draw[-] (k3) to (k4);

\draw (k1) -- (N1);

\end{scope}
\end{tikzpicture}};
		\node[align=justify, font=\scriptsize] at (0.5,0.05) {H. W.};
		\node[align=justify, font=\small] at (6.2,-0.4) {$(2,2,2,2)$};
		\draw[->, -stealth,  line width=0.3em](0.15,-0.35) -- (0.95,-0.35);
		\node at (3.5,0) {\begin{tikzpicture}[baseline,font=\footnotesize]
 \begin{scope}[auto, every node/.style={minimum size=0.65cm}]
\def \spac {0.9cm}

\node[circle, draw](k1) at (0,0) {$4$};
\node[circle, draw](k2) at (1*\spac,0) {$6$};
\node[circle, draw](k3) at (2*\spac,0*\spac) {$6$};
\node[circle, draw](k4) at (3*\spac,0*\spac) {$4$};

\node[draw, inner sep=0.1cm,minimum size=0.55cm](N1) at (0*\spac,1*\spac) {$2$};
\node[draw, inner sep=0.1cm,minimum size=0.55cm](N2) at (1*\spac,1*\spac) {$2$};
\node[draw, inner sep=0.1cm,minimum size=0.55cm](N3) at (2*\spac,1*\spac) {$2$};
\node[draw, inner sep=0.1cm,minimum size=0.55cm](N4) at (3*\spac,1*\spac) {$2$};

\draw[-] (k1) to (k2);
\draw[-] (k2) to (k3);
\draw[-] (k3) to (k4);

\draw (k1) -- (N1);
\draw (k2) -- (N2);
\draw (k3) -- (N3);
\draw (k4) -- (N4);

\end{scope}
\end{tikzpicture}};
		\end{tikzpicture}
	\end{center}
	\caption{An example of how one symmetrizes a little string quiver of $A_n$ using Hanany-Witten transitions, to end up with a weighted Dynkin diagram. The Coulomb parameters in red are frozen, and therefore do not increase the Coulomb branch dimension.  In this example, no matter what the details of the transition are, the resulting symmetric quiver is always (2,2,2,2), the full puncture. Note some of the masses are equal to each other  in the resulting quiver, as they should after Higgs flow.}
	\label{fig:ExampleWeightedDynkin}
\end{figure}
\FloatBarrier
Then, it turns out that all $A_n$ little string quivers that are not symmetric under $\mathbb{Z}_2$ reflection can be uniquely written after Higgs flow as weighted Dynkin diagrams with correct Bala--Carter label. For instance, one can show that the full puncture of $A_n$, with Bala--Carter label $\varnothing$, can be symmetrized uniquely to give the weighted Dynkin diagram $(2,2,\ldots,2,2)$. See figure \ref{fig:ExampleWeightedDynkin}.\\

This map between little string quivers and weighted Dynkin diagrams is one-to-one for $\fg=A_n$, but many-to-one for the other algebras, as a large number of different little string quivers typically describe one and the same defect in those cases. Nevertheless, the map always exists.\\

We now come to another result about weighted Dynkin diagrams, motivated by their apparent connection to little string defects we have pointed out: the dimension of a nilpotent orbit can be easily computed from its weighted Dynkin diagram.

\subsection{Dimension Formula}
\label{sec:wformula}
Recall that the ``flavor symmetry rank'' of a weighted Dynkin diagram never exceeds 2 (as the flavor symmetry is always a product of $U(1)$ and $U(2)$  groups only). This is a claim about the hypermultiplets of the quiver theory. There exists a ``vector multiplet'' counterpart to this statement, which is given by the following mathematical statement:\\

We interpret the weighted Dynkin diagram of a nilpotent orbit $\cO$ as a weight $\omega$, written down in the Dynkin basis. We then compute the sum of the inner products of all the positive roots of $\fg$ with this weight. This gives a vector of non-negative integers. Truncating the entries of this vector at 2 and taking the sum of the entries gives the (real) dimension of $\cO$.\\

This result can be derived from the following dimension formula for nilpotent orbits\footnote{We thank Axel Kleinschmidt for pointing out this proof to us.} (cf.\ for instance \cite{Collingwood:1993}):

\begin{equation}
\label{eq:dimform}
\dim \cO=\dim \fg- \dim \fg_0-\dim \fg_1,
\end{equation}
where 
\begin{equation}
\label{eq:grading}
\fg_i=\{Z\in\fg|[H,Z]=i\cdot Z\},
\end{equation}
and where $H$ is the semisimple element in the $\mathfrak{sl}_2$ triple corresponding to $\cO$.

Note that whenever $Z\in \fg_\beta$ for a root $\beta$, $[H,Z]=\beta(H)Z$. So 
$$
\fg_i=\bigoplus_{\substack{\beta\in\Phi,\\\beta(H)=i}}\fg_\beta.
$$
On the other hand, if $\fg$ is simply laced, then the inner product of the weighted Dynkin diagram weight $\omega$ with a root $\beta$ is just
$$
 \left\langle\sum_{i=1}^n\alpha_i(H)\omega_i,\beta\right\rangle=\sum_{i=1}^n \alpha_i(H)\langle\omega_i,\beta\rangle=\beta(H),
$$
where $\alpha_i$ and $\omega_i$ are the simple roots and fundamental weights of $\fg$, respectively.

Thus, if $\fg$ is simply laced, the above inner products just give us the grading \ref{eq:grading}.

The prescription we give is therefore equivalent to the dimension formula  \ref{eq:dimform}; namely, 

\begin{equation}
\begin{split}
\dim(\fg_1)+2\sum_{i\geq 2}\dim(\fg_i)&=\dim(\fg_1)+\sum_{i\geq 2}\dim(\fg_i)+\sum_{i\leq -2}\dim(\fg_i)\\
&=\dim(\fg_1)+\dim(\fg)-\sum_{-1\leq i\leq1}\dim(\fg_i)\\
&=\dim(\fg_1)+\dim(\fg)-2\dim(\fg_1)-\dim(\fg_0)\\
&=\dim(\fg)-\dim(\fg_0)-\dim(\fg_1).
\end{split}
\end{equation}

\begin{example} Let us take the example of the weighted Dynkin diagram (2,1,1,2) in the algebra $\fg=A_4$. We write $\omega=[2,1,1,2]$ as a weight in Dynkin basis. 

The positive roots $\Phi^+$ of $A_4$ are
\begin{equation*}
\begin{split}
(h_1-h_5,\;h_2-h_5,\;h_1-h_4,\;h_2-h_4,\;h_3-h_5,\;h_1-h_3,\;h_2-h_3,\;h_3-h_4,\;h_4-h_5,\;h_1-h_2)
\end{split}
\end{equation*}
Calculating the inner product of all of these positive roots with $\omega$ gives the numbers
\begin{align*}
\langle \Phi^+,\omega\rangle=(6,4,4,2,3,3,1,1,2,2).
\end{align*}
Truncating at multiplicity 2, the sum of the inner products is $2 \times 8 + 1 \times 2 = 18$, which is indeed the dimension of the nilpotent orbit denoted by the diagram $(2,1,1,2)$.\\
\end{example}

Note that the theories $T^{2d}$ we have been studying can be interpreted as 3d $\cN=4$ theories. It is then interesting to compare this formula to the dimension of the Coulomb branch of a 3d $\cN=4$ quiver theory \cite{Bullimore:2015lsa}, which is given by a slice in the affine Grassmannian \cite{Braverman:2016pwk}. In that setup, the dimension can be calculated by the exact same procedure, coming from a monopole formula \cite{Moore:2014jfa}, but without truncating the inner products at the value 2. For conformal theories, this is simply the sum of the ranks of the gauge groups. \\

Lastly, we want to emphasize that the above formula we gave does not compute the Coulomb branch dimension of the defect theory $T^{2d}_{m_s\rightarrow\infty}$ denoted by the weighted Dynkin diagram. Instead, the Coulomb branch dimension is given by the dimension of the diagram's image under the Spaltenstein map. Note that not all nilpotent orbits are in the image of the Spaltenstein map, so in many cases, it is unclear what the physical interpretation of the dimension formula should be.

\section*{Acknowledgments}
We want to thank Mina Aganagic, Tudor Dimofte, Jacques Distler, Ori Ganor, Amihay Hanany, Alex Takeda, Yuji Tachikawa and Michael Viscardi for helpful discussions and comments.
Furthermore, we want thank Axel Kleinschmidt for explaining the derivation of the dimension formula in section \ref{sec:wformula} to us.

The research of N. H. and C. S. is supported in part by the Berkeley Center for Theoretical
Physics,  by  the  National  Science  Foundation  (award PHY-1521446) and by the
US Department of Energy under Contract DE-AC02-05CH11231.
\appendix
\begin{landscape}
\pagestyle{plain}
\section{$E_n$ little string defects}
\label{sec:results}

As an application of the Bala--Carter classification, we now present a table of the defects of the $E_n$ little string. Unpolarized defects are shaded in yellow. For each defect type, we give a set $\cW_{\cS}$ of weights, along with the low energy 2d quiver gauge theory $T^{2d}$ on the D3 branes that results from it. The Bala--Carter label that designates the nilpotent orbit in the CFT limit $m_s\rightarrow\infty$ is written in the left column. Each set $\cW_{\cS}$ is a distinguished set, in the sense of section \ref{sec:review}; in particular, the weights $\omega_i$ of $\cW_{\cS}$ satisfy
\[
\vec\beta\cdot\vec\alpha_i=0 \quad \forall \vec\alpha_i\in\Theta,
\]
with $\vec\beta=\sum_{i=1}^{|\cW_{\cS}|}\beta_i\; \omega_i$. This constraint has an interpretation as a level 1 null state condition of $\fg$-Toda. For unpolarized defects, a subscript is added to the weights, specifying the representation they are taken in. This corresponds to giving the additional simple root label $a_i$ in the Bala--Carter picture. For polarized defects, no subscript is needed for the weights.\\

The dual orbit is the orbit describing the Coulomb branch of $T^{2d}_{m_s\rightarrow \infty}$; for polarized defects, this is given by the Spaltenstein dual of the Bala--Carter label. For unpolarized defects, these dual orbits had to be conjectured based on other approaches, such as dimension counting.
The dimension of this dual orbit describing the Coulomb branch is given by $d$.\\

Note the quivers are either literally the weighted Dynkin diagrams as given in the  literature, or are quivers that can be made to be weighted Dynkin diagrams after Higgs flow.\\

\begin{longtable}{llb{4cm}cll}
\caption{Results for $E_6$}\\
Orbit&$\Theta$&Weights&Quiver&Dual orbit&$d$\\
\hline\endhead\\
$0$ & $\varnothing$ &\scriptsize$[\phantom{-}0,\phantom{-}0,\phantom{-}0,\phantom{-}0,\phantom{-}0,-1]$\newline$[-1,\phantom{-}0,\phantom{-}0,\phantom{-}0,-1,\phantom{-}1]$\newline$[\phantom{-}1,\phantom{-}0,\phantom{-}0,-1,\phantom{-}0,\phantom{-}1]$\newline$[\phantom{-}0,-1,\phantom{-}0,\phantom{-}0,\phantom{-}1,\phantom{-}1]$\newline$[\phantom{-}0,\phantom{-}1,-1,\phantom{-}1,\phantom{-}0,\phantom{-}0]$\newline$[\phantom{-}0,\phantom{-}0,\phantom{-}1,\phantom{-}0,\phantom{-}0,-1]$\newline$[\phantom{-}0,\phantom{-}0,\phantom{-}0,\phantom{-}0,\phantom{-}0,-1]$ & \includegraphics{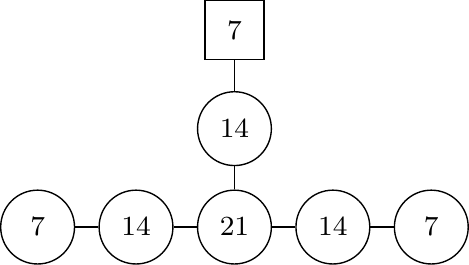} & $E_6$ & 72\\$A_1$ & $\{\alpha_1\}$ &\scriptsize$[\phantom{-}0,\phantom{-}0,\phantom{-}0,\phantom{-}0,-1,\phantom{-}0]$\newline$[\phantom{-}0,\phantom{-}0,\phantom{-}0,-1,\phantom{-}1,\phantom{-}0]$\newline$[\phantom{-}0,\phantom{-}1,\phantom{-}0,-1,\phantom{-}0,\phantom{-}0]$\newline$[\phantom{-}0,\phantom{-}0,\phantom{-}0,\phantom{-}1,\phantom{-}0,-1]$\newline$[\phantom{-}0,\phantom{-}0,-1,\phantom{-}1,\phantom{-}0,\phantom{-}1]$\newline$[\phantom{-}0,-1,\phantom{-}1,\phantom{-}0,\phantom{-}0,\phantom{-}0]$ & \includegraphics{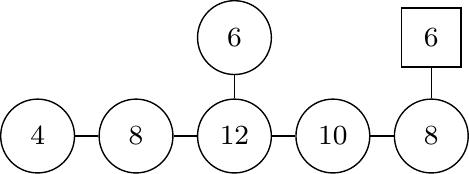} & $E_6(a_1)$ & 70\\$2A_1$ & $\{\alpha_1,\alpha_3\}$ &\scriptsize$[\phantom{-}0,\phantom{-}0,\phantom{-}0,\phantom{-}1,-1,-1]$\newline$[\phantom{-}0,\phantom{-}0,\phantom{-}0,\phantom{-}0,\phantom{-}1,-1]$\newline$[\phantom{-}0,-1,\phantom{-}0,\phantom{-}0,\phantom{-}0,\phantom{-}1]$\newline$[\phantom{-}0,\phantom{-}1,\phantom{-}0,-1,\phantom{-}0,\phantom{-}0]$\newline$[\phantom{-}0,\phantom{-}0,\phantom{-}0,\phantom{-}0,\phantom{-}0,\phantom{-}1]$ & \includegraphics{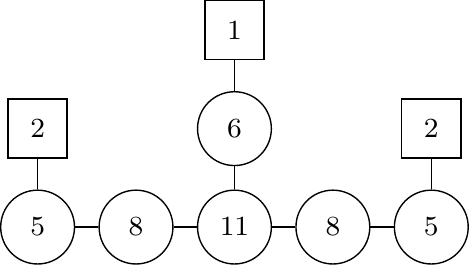} & $D_5$ & 68\\
\end{longtable}
\clearpage
\begin{longtable}{llb{4.2cm}cll}
\caption{Results for $E_7$}\\
Orbit&$\Theta$&Weights&Quiver&Dual orbit&$d$\\
\hline\endhead\\
$0$ & $\varnothing$ &\scriptsize$[-1,\phantom{-}0,\phantom{-}0,\phantom{-}0,\phantom{-}0,\phantom{-}0,\phantom{-}0]$\newline$[\phantom{-}1,\phantom{-}0,\phantom{-}0,-1,\phantom{-}0,\phantom{-}1,\phantom{-}0]$\newline$[\phantom{-}0,-1,\phantom{-}1,\phantom{-}0,\phantom{-}0,-1,\phantom{-}0]$\newline$[-1,\phantom{-}1,\phantom{-}0,\phantom{-}0,\phantom{-}0,-1,\phantom{-}0]$\newline$[\phantom{-}0,\phantom{-}0,\phantom{-}0,\phantom{-}0,\phantom{-}0,\phantom{-}1,\phantom{-}0]$\newline$[\phantom{-}0,\phantom{-}0,-1,\phantom{-}1,\phantom{-}0,\phantom{-}0,\phantom{-}1]$\newline$[\phantom{-}1,\phantom{-}0,-1,\phantom{-}1,-1,\phantom{-}0,\phantom{-}0]$\newline$[\phantom{-}0,\phantom{-}0,\phantom{-}1,-1,\phantom{-}1,\phantom{-}0,-1]$ & \includegraphics{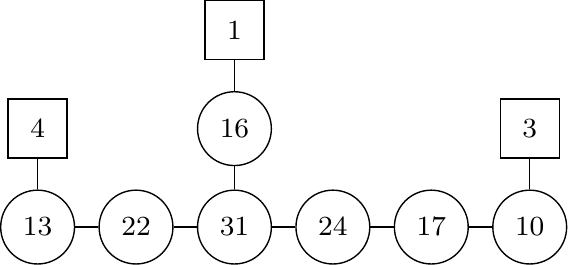} & $E_7$ & 126\\$A_1$ & $\{\alpha_5\}$ &\scriptsize$[-1,\phantom{-}0,\phantom{-}0,\phantom{-}0,\phantom{-}0,\phantom{-}0,\phantom{-}0]$\newline$[\phantom{-}1,\phantom{-}0,\phantom{-}0,-1,\phantom{-}0,\phantom{-}1,\phantom{-}0]$\newline$[\phantom{-}0,-1,\phantom{-}1,\phantom{-}0,\phantom{-}0,-1,\phantom{-}0]$\newline$[-1,\phantom{-}1,\phantom{-}0,\phantom{-}0,\phantom{-}0,-1,\phantom{-}0]$\newline$[\phantom{-}0,\phantom{-}0,\phantom{-}0,\phantom{-}0,\phantom{-}0,\phantom{-}1,\phantom{-}0]$\newline$[\phantom{-}0,\phantom{-}0,-1,\phantom{-}1,\phantom{-}0,\phantom{-}0,\phantom{-}1]$\newline$[\phantom{-}1,\phantom{-}0,\phantom{-}0,\phantom{-}0,\phantom{-}0,\phantom{-}0,-1]$ & \includegraphics{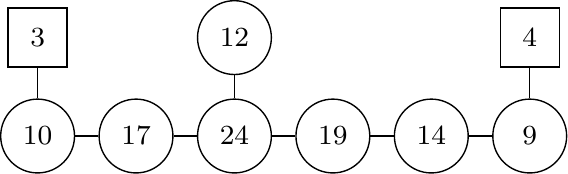} & $E_7(a_1)$ & 124\\
\end{longtable}
\clearpage
\begin{longtable}{llb{4.5cm}cll}
\caption{Results for $E_8$}\\
Orbit&$\Theta$&Weights&Quiver&Dual orbit&$d$\\
\hline\endhead\\
$0$ & $\varnothing$ &\scriptsize$[-1,\phantom{-}0,\phantom{-}0,\phantom{-}0,\phantom{-}0,\phantom{-}0,\phantom{-}0,\phantom{-}0]$\newline$[\phantom{-}1,-1,\phantom{-}1,-1,\phantom{-}0,\phantom{-}0,\phantom{-}0,\phantom{-}0]$\newline$[\phantom{-}0,\phantom{-}1,-1,\phantom{-}1,\phantom{-}0,-1,\phantom{-}0,\phantom{-}0]$\newline$[-1,-1,\phantom{-}1,-1,\phantom{-}0,\phantom{-}1,-1,\phantom{-}0]$\newline$[\phantom{-}1,\phantom{-}2,-2,\phantom{-}1,\phantom{-}0,\phantom{-}0,\phantom{-}1,\phantom{-}0]$\newline$[-1,\phantom{-}0,\phantom{-}0,\phantom{-}1,-1,\phantom{-}0,\phantom{-}0,-1]$\newline$[\phantom{-}1,-1,\phantom{-}0,\phantom{-}0,\phantom{-}1,\phantom{-}0,\phantom{-}0,\phantom{-}1]$\newline$[\phantom{-}0,\phantom{-}0,\phantom{-}2,-2,\phantom{-}0,\phantom{-}0,\phantom{-}0,-1]$\newline$[\phantom{-}0,\phantom{-}0,-1,\phantom{-}1,\phantom{-}0,\phantom{-}0,\phantom{-}0,\phantom{-}1]$ & \includegraphics{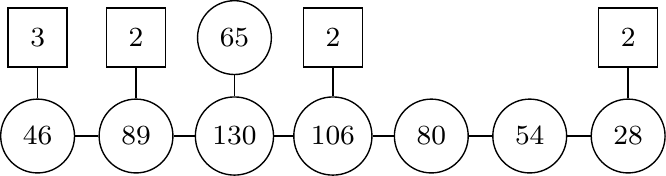} & $E_8$ & 240\\$A_1$ & $\{\alpha_7\}$ &\scriptsize$[-1,\phantom{-}0,\phantom{-}0,\phantom{-}0,\phantom{-}0,\phantom{-}0,\phantom{-}0,\phantom{-}0]$\newline$[\phantom{-}1,-1,\phantom{-}1,-1,\phantom{-}0,\phantom{-}0,\phantom{-}0,\phantom{-}0]$\newline$[\phantom{-}0,\phantom{-}1,-1,\phantom{-}1,\phantom{-}0,-1,\phantom{-}0,\phantom{-}0]$\newline$[\phantom{-}0,\phantom{-}1,-1,\phantom{-}0,\phantom{-}0,\phantom{-}1,\phantom{-}0,\phantom{-}0]$\newline$[-1,\phantom{-}0,\phantom{-}0,\phantom{-}1,-1,\phantom{-}0,\phantom{-}0,-1]$\newline$[\phantom{-}1,-1,\phantom{-}0,\phantom{-}0,\phantom{-}1,\phantom{-}0,\phantom{-}0,\phantom{-}1]$\newline$[\phantom{-}0,\phantom{-}0,\phantom{-}2,-2,\phantom{-}0,\phantom{-}0,\phantom{-}0,-1]$\newline$[\phantom{-}0,\phantom{-}0,-1,\phantom{-}1,\phantom{-}0,\phantom{-}0,\phantom{-}0,\phantom{-}1]$ & \includegraphics{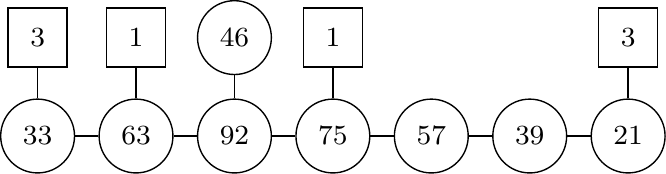} & $E_8(a_1)$ & 238\\
\end{longtable}
\end{landscape}

\section{Zero Weight Multiplicity}
\label{sec:zero}
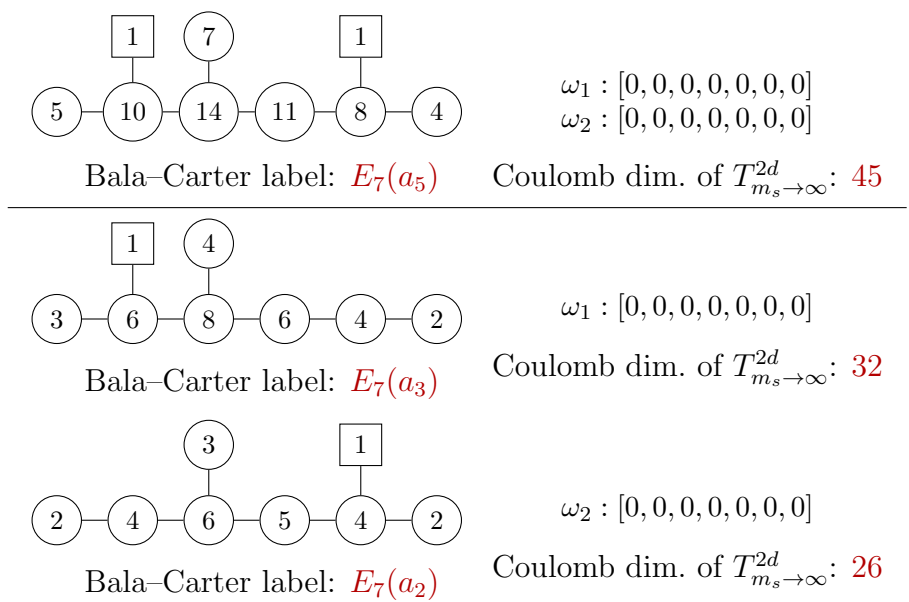
\begin{figure}[htpb]
	\begin{center}
	\begin{tabular}{c}
\begin{tikzpicture}[baseline,font=\small]
	\node at (-2.5,0) {\begin{tikzpicture}[baseline,font=\footnotesize]
 \begin{scope}[auto, every node/.style={minimum size=0.65cm}]
\def \spac {1.0cm}

\node[circle, draw](k1) at (0,0) {$5$};
\node[circle, draw](k2) at (1*\spac,0) {$10$};
\node[circle, draw](k3) at (2*\spac,0*\spac) {$14$};
\node[circle, draw](k4) at (3*\spac,0*\spac) {$11$};
\node[circle, draw](k5) at (4*\spac,0*\spac) {$8$};
\node[circle, draw](k6) at (5*\spac,0*\spac) {$4$};
\node[circle, draw](k7) at (2*\spac,1*\spac) {$7$};

\node[draw, inner sep=0.1cm,minimum size=0.55cm](N2) at (1*\spac,1*\spac) {$1$};
\node[draw, inner sep=0.1cm,minimum size=0.55cm](N5) at (4*\spac,1*\spac) {$1$};

\draw[-] (k1) to (k2);
\draw[-] (k2) to (k3);
\draw[-] (k3) to (k4);
\draw[-] (k4) to (k5);
\draw[-] (k5) to (k6);
\draw[-] (k3) to (k7);

\draw (k2) -- (N2);
\draw (k5) -- (N5);

\end{scope}
\end{tikzpicture}};
		\node[align=justify, font=\small] at (3.3,-0.35) {$\omega_1: [0,0,0,0,0,0,0]$\\$\omega_2: [0,0,0,0,0,0,0]$};
		\node[align=justify, font=\normalsize] at (3.3,-1.35) {Coulomb dim.\ of $T^{2d}_{m_s\to\infty}$: {\color{red!70!black}45}};
		\node[align=justify, font=\normalsize] at (-2.3,-1.35) {Bala--Carter label: {\color{red!70!black}$E_7(a_5)$}};
		\end{tikzpicture}\\
		\hline
		\begin{tikzpicture}[baseline,font=\small]
	\node at (-2.5,0) {\begin{tikzpicture}[baseline,font=\footnotesize]
 \begin{scope}[auto, every node/.style={minimum size=0.65cm}]
\def \spac {1.0cm}

\node[circle, draw](k1) at (0,0) {$3$};
\node[circle, draw](k2) at (1*\spac,0) {$6$};
\node[circle, draw](k3) at (2*\spac,0*\spac) {$8$};
\node[circle, draw](k4) at (3*\spac,0*\spac) {$6$};
\node[circle, draw](k5) at (4*\spac,0*\spac) {$4$};
\node[circle, draw](k6) at (5*\spac,0*\spac) {$2$};
\node[circle, draw](k7) at (2*\spac,1*\spac) {$4$};

\node[draw, inner sep=0.1cm,minimum size=0.55cm](N2) at (1*\spac,1*\spac) {$1$};

\draw[-] (k1) to (k2);
\draw[-] (k2) to (k3);
\draw[-] (k3) to (k4);
\draw[-] (k4) to (k5);
\draw[-] (k5) to (k6);
\draw[-] (k3) to (k7);

\draw (k2) -- (N2);

\end{scope}
\end{tikzpicture}};
		\node[align=justify, font=\small] at (3.3,-0.35) {$\omega_1: [0,0,0,0,0,0,0]$};
		\node[align=justify, font=\normalsize] at (-2.3,-1.35) {Bala--Carter label: {\color{red!70!black}$E_7(a_3)$}};
		\node[align=justify, font=\normalsize] at (3.3,-1.15) {Coulomb dim.\ of $T^{2d}_{m_s\to\infty}$: {\color{red!70!black}32}};
		\end{tikzpicture}\\
		\begin{tikzpicture}[baseline,font=\small]
	\node at (-2.5,0) {\begin{tikzpicture}[baseline,font=\footnotesize]
 \begin{scope}[auto, every node/.style={minimum size=0.65cm}]
\def \spac {1.0cm}

\node[circle, draw](k1) at (0,0) {$2$};
\node[circle, draw](k2) at (1*\spac,0) {$4$};
\node[circle, draw](k3) at (2*\spac,0*\spac) {$6$};
\node[circle, draw](k4) at (3*\spac,0*\spac) {$5$};
\node[circle, draw](k5) at (4*\spac,0*\spac) {$4$};
\node[circle, draw](k6) at (5*\spac,0*\spac) {$2$};
\node[circle, draw](k7) at (2*\spac,1*\spac) {$3$};

\node[draw, inner sep=0.1cm,minimum size=0.55cm](N5) at (4*\spac,1*\spac) {$1$};

\draw[-] (k1) to (k2);
\draw[-] (k2) to (k3);
\draw[-] (k3) to (k4);
\draw[-] (k4) to (k5);
\draw[-] (k5) to (k6);
\draw[-] (k3) to (k7);

\draw (k5) -- (N5);

\end{scope}
\end{tikzpicture}};
		\node[align=justify, font=\small] at (3.3,-0.35) {$\omega_2: [0,0,0,0,0,0,0]$};
		\node[align=justify, font=\normalsize] at (-2.3,-1.35) {Bala--Carter label: {\color{red!70!black}$E_7(a_2)$}};
		\node[align=justify, font=\normalsize] at (3.3,-1.15) {Coulomb dim.\ of $T^{2d}_{m_s\to\infty}$: {\color{red!70!black}26}};
		\end{tikzpicture}\\
	\end{tabular}
	\end{center}
	\caption{In the little string, at finite $m_s$, defects add up in a linear fashion. For instance, the $E_7$ defect shown on top is the sum of the two defects shown under it. For polarized defects, we usually refer to this situation as a four-punctured sphere: two full punctures and two punctures each labeled by the zero weight. However, in the $m_s\rightarrow\infty$, the defect really should be thought of as a three punctured sphere, with two full punctures and an exotic puncture given by a combination of the two zero weights, which cannot be split apart. As a quick check, this is confirmed by noting that the Coulomb branch dimension of $T^{2d}_{m_s\rightarrow\infty}$ is not additive.}
	\label{fig:e7add}
\end{figure}

We make a comment about the multiplicity of the zero weight in unpolarized defects. This is relevant for two of the defects analyzed in appendix \ref{sec:results}: one has Bala--Carter label $E_7(a_5)$ in $\fg=E_7$, and the other has Bala--Carter label $E_8(b_5)$ in $\fg=E_8$. For both of these, $\cW_{\cS}$ is the set of the zero weight only, but appearing \emph{twice}. In the little string, at finite $m_s$, defects usually add up in a linear fashion \cite{Aganagic:2014oia, Aganagic:2015cta}. If a subset of weights in $\cW_{\cS}$ adds up to zero, then one is simply describing more than one elementary defect. In the case of polarized defects, where a direct Toda interpretation is available, we would refer to this situation as a higher-than-three point function on the sphere. We note here that for the two unpolarized defects we mentioned, this is not the case. For both cases, the zero weight is required to appear twice and does characterize a single exotic defect, with  Bala--Carter label given above. In particular, $E_7(a_5)$ and $E_8(b_5)$ are not engineered in the little string as the sum of two elementary defects with a single zero weight. See Figure \ref{fig:e7add} for the example of $E_7(a_5)$.

\mciteSetMidEndSepPunct{}{}{} 
\bibliography{summary}
\bibliographystyle{utphysmcite}

\end{document}